# Theory of field-modulated spin-valley-orbital pseudospin physics

Feng-Wu Chen[1] and Yu-Shu G. Wu[1, 2, †]

[1] *Department of Physics, National Tsing-Hua University, Hsin-Chu 30013, Taiwan, ROC*

[2] *Department of Electrical Engineering, National Tsing-Hua University, Hsin-Chu 30013, Taiwan, ROC*

Pioneering studies in transition metal dichalcogenides have demonstrated convincingly the co-existence of multiple angular momentum degrees of freedom – of spin (1/2 $s_z = \pm 1/2$), valley ($\tau$ = K, K' or $\pm 1$), and atomic orbital ($l_z = \pm 2$) origins – in the valence band with strong interlocking among them, which results in noise-resilient pseudospin states ideal for spintronic type applications. With field modulation a powerful, universal means in physics studies and applications, this work develops, from bare models in the context of complicated band structure, a general effective theory of field-modulated spin-valley-orbital pseudospin physics that is able to describe both intra- and inter- valley dynamics. Based on the theory, it predicts and discusses the linear response of a pseudospin to external fields of arbitrary orientations. Paradigm field configurations are identified for pseudospin control including pseudospin flipping. For a nontrivial example, it presents a spin-valley-orbital quantum computing proposal, where the theory is applied to address all-electrical, simultaneous control of $s_z$, $\tau$, and $l_z$ for qubit manipulation. It demonstrates the viability of such control with static field effects and an additional dynamic electric field. An optimized qubit manipulation time ~ O($ns$) is given.

## I. INTRODUCTION

The discovery of spin degree of freedom (DoF) in the Stern-Gerlach experiment has opened up a new era in quantum physics. Striking spin phenomena include spin Hall effect [1-5] and spin-dependent transport such as giant [6,7] or colossal [8,9] magnetoresistance, to name a few. Effective field modulation with Rashba [10] or Zeeman [11,12] effects plays a crucial role in pioneering studies and device proposals, including, for example, spin FETs, [13] spin quantum computing [14,15], and so on, in the category of spintronics. [16,17]

With the rise of 2D materials [18-20] recent years have seen a rapid expansion of research from spin to angular momenta on various length scales. Notably, in 2D crystals of hexagonal symmetry, "valley pseudospin" - a binary electron DoF has emerged, which derives from the existence of doubly degenerate, time-reversal-conjugated energy band valleys at Dirac corners (K and K′) of Brillouin zone. [21-23] Exotic topological transport phenomena arise due to the valley DoF, such as valley Hall effect [21,23,24] in graphene [25-27] and transition metal dichalcogenides (TMDCs) [28]. In these materials, electron "valley" magnetic moments or angular momenta [21,29] are manifested on the unit-cell scaled orbital motion, and can interact with an in-plane electric field in the form

$$H_{VOI} \propto \left( \vec{k} \times \vec{\varepsilon}_{//} \right) \cdot \vec{\mu}_\tau \quad (1)$$

known as valley-orbit interaction (VOI) ($\vec{k}$ = in-plane electron wave vector; $\vec{\varepsilon}_{//}$ = in-plane electric field; and $\vec{\mu}_\tau$ = valley magnetic moment). [30,31] Such interaction is similar to the spin-orbit interaction (SOI) and constitutes a useful mechanism for applications in the category of valleytronics.

Among 2D materials, TMDCs stand out as a unique family characterized by the presence of strong SOI and plural angular momentum DoFs – of spin, valley, and atomic orbital origins. Pioneering studies [32-35] have convincingly demonstrated the existence of rich quantum physics in TMDCs from intriguing interplay among co-existing DoFs and SOI. With TMDCs, the spectrum of spintronic type physics is broadened for varied applications. **Figure 1** summarizes important elements in single-particle, spintronic type physics in solids, in the four categories: spin, valley, spin-valley, and spin-valley-orbital (SVO), with the variety summarized here hosting a vast range of possibilities, including all- spintronic and valleytronic circuits. The figure places an emphasis on field control or modulation of the physics. In general, electrical fields, as well as magnetic fields in vertical [30, 40-42] or in-plane directions [30, 43] can be introduced and coupled to the various magnetic moments (or angular momenta), in order to tune the physics.

**Figures 1(a)** and **1(b)** illustrate the modulation of electrical nature via SOI and VOI mechanisms, respectively, showing a similarity between the two, namely, that the presence of an electric field results in an effective magnetic field ($\vec{B}_{eff}$) and a corresponding interaction with the magnetic moment. On the other hand, the similarity exhibited is superficial, since the two mechanisms differ fundamentally in physics: SOI has a relativistic origin, whereas VOI is a pseudo-relativistic effect determined by the band-structure physics. In addition, while in the SOI case both the spin magnetic moment ($\vec{\mu}_s$) and electric field ($\varepsilon$) can be arbitrarily oriented, in the VOI case the valley magnetic moment ($\vec{\mu}_\tau$) derives from the circulating current inside each hexagon of the honeycomb lattice and, thus, always points out of plane (// $\hat{z}$), which constrains the corresponding $\varepsilon$ ($\vec{B}_{eff}$) to be in-plane (out-of-plane), e.g., $\varepsilon = \vec{\varepsilon}_{//}$, making VOI a valley index-conserved interaction. Overall, the availability of and the flexibility in modulation via SOI or VOI have profound implications for industrial applications, e.g., electrical gate-controlled ICs.

Apart from the control, another critical issue – state coherence faces spintronic type applications. Generally speaking, robust state coherence is required for applications in a noisy setting, in particular those at the room temperature. In connection with this respect, as well as for applications in general, TMDCs exhibit the following band structure features with important implications. [29,39-42,44-47] In the monolayer case, they have a unit cell consisting of one transition metal atom (M) and two chalcogen ones ($X_2$), a semiconductor band structure with direct band gap (1-2 $eV$) at

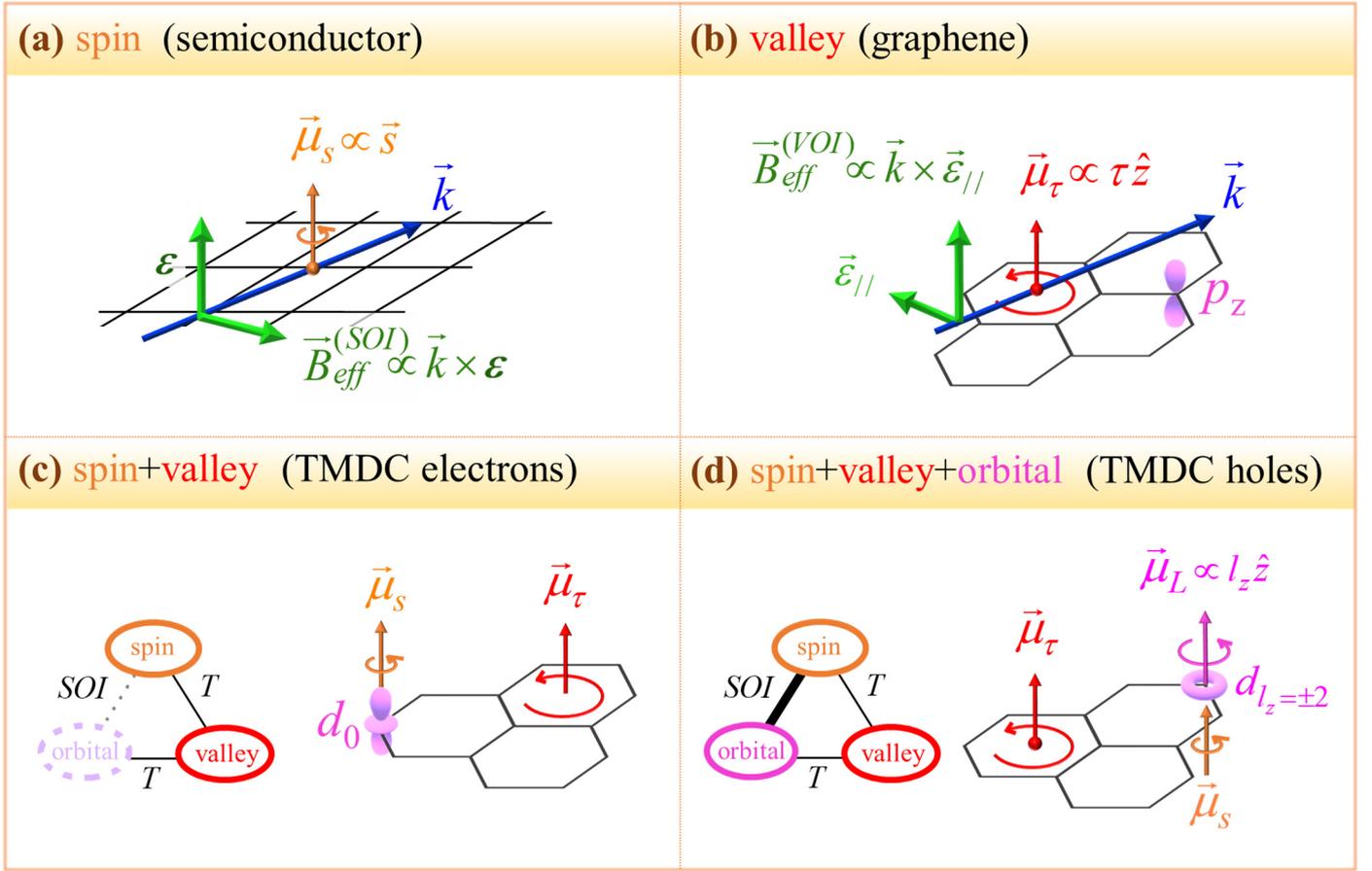

**Figure 1** The spintronic type physics in solids, in the four categories: **(a)** spin, **(b)** valley, **(c)** spin-valley, and **(c)** spin-valley-orbital physics, with magnetic moments $\vec{\mu}_s$, $\vec{\mu}_\tau (\tau = K, K' \text{ or } 1, -1)$, and $\vec{\mu}_L$ in association with spin, valley, and atomic orbital DoFs, respectively. **(a)** and **(b)** show a unified methodology for electrical manipulation of electrons, based on the interaction between a pseudo-magnetic field $\vec{B}_{eff} \propto \vec{k} \times \vec{\varepsilon}$ ( $\vec{\varepsilon}$ = static electric field and $\vec{k}$ = electron wave vector) and magnetic moments, e.g., $\vec{\mu}_s$ in **(a)** and $\vec{\mu}_\tau$ in **(b)**, with the interaction mechanism being SOI in **(a)** for semiconductors such as InAs [36], InSb [37], InGaAs [38], and etc., and VOI in **(b)** for $2p_z$ electrons in graphene. In **(c),** for conduction band electrons in TMDCs, spin and valley DoFs co-exist, but the atomic orbital DoF is basically frozen at $d_0$ with feeble components of $d_{\pm 1}$ and $p_{\pm 1}$, [39] which induce a weak $\vec{\mu}_L$ as well as SOI coupling between spin and atomic orbital DoFs ( $\propto \vec{\mu}_s \cdot \vec{\mu}_L$ ). In **(d)** for valence band holes in TMDCs, spin, valley, and atomic orbital ($d_{\pm 2}$) DoFs co-exist, with a strong SOI-induced coupling between spin and atomic orbital. In both **(c)** and **(d)**, a conjugated relation due to the time-reversal symmetry (denoted by "T" in the figure) exists among spin, valley, and atomic orbital DoFs, where $\vec{\mu}_s$ and $\vec{\mu}_L$ are flipped for degenerate electron states opposite in $\tau$.

Dirac points, and valence (conduction) band edge states primarily derived from the $d_{\pm 2}$ ($d_0$) orbital of M. Due to the SOI in M, spin-orbit splitting occurs at band edges, with the splitting much more pronounced in the valence band (0.1-0.5 $eV$) than in the conduction band (3-50 $meV$). The existence of band gap makes it possible to create electric gate-defined confining structures, e.g., quantum dots [48-51] or wires [52] useful for general applications.

**Figures 1(c)** and **1(d)** summarize the implications of foregoing band structure features for pseudospin physics in TMDCs. They show the coupling among spin, valley, and atomic orbital DoFs, in the conduction and valence bands, respectively. Due to such coupling, novel pseudospin states emerge near the gap, as experimentally confirmed by the generation of valley polarization with optical excitations [32-34]. Notably, as shown in **Figure 1(c)**, since spin and valley in the conduction band are only weakly SOI-coupled, they can be used nearly independently and simultaneously.[41] Such advantage has recently been exploited, resulting in unique spin-valley quantum computing proposals[53-55] and versatile electron qubit schemes.[43, 56]

On the other hand, as indicated in **Figure 1(d)**, a distinct type of pseudospin physics exists in the valence band. At the valence band maximum (*VBM*), a Kramers pair of states, denoted as $|K\rangle$ (or $|VBM, K\rangle$) and $|K'\rangle$ (or $|VBM, K'\rangle$) throughout the work, are formed at K and K' and characterized by opposite values of quantum indices, ($1/2\ s_z = 1/2$, $\tau = 1$ or K, $l_z = 2$) and ($1/2\ s_z = -1/2$, $\tau = -1$ or K', $l_z = -2$), respectively, where $s_z$, $\tau$, and $l_z$ refer to spin, valley, and atomic orbital indices of the electron, respectively. Such pair of states define a unique "spin-valley-orbital pseudospin", extremely noise-resilient due to strong SOI-induced interlocking among $s_z$, $\tau$, and $l_z$ against individual index fluctuations.[29] Experimentally[57-59] and theoretically,[60] the valley lifetime of holes is reported to be enhanced over that of electrons by 10-100 times reaching O(10 $\mu s$) at 5-10 *K*. Such advantage fosters quite an exciting promise for pseudospin-based studies, applications at low temperatures such as quantum computing, and also room-temperature devices such as pseudospin filters and FETs, and has motivated researchers from a wide range of disciplines.

This work searches a theory for spin-valley-orbital pseudospin physics studies and applications. Concerning the latter, the following nontrivial issue is to be addressed, namely, while the pseudospin coherence is a key advantage, the underlying mechanism for coherence - sturdy interlocking among existing DoFs - also poses a tremendous challenge to the control of pseudospins, especially in the case of pseudospin flip manipulation. In view of such issue, this work proceeds as summarized in the following. Overall, it formulates a general theoretical framework for the pseudospin physics in external fields, in the context of complicated TMDC band structure. It starts by setting up multi-band "bare models", which account for effects of elastic valley-flip scattering due to impurities in the bulk or boundaries of quantum structures. Inclusion of such scattering, when combined with that of spin- and atomic orbital- mixing mechanisms as well as field effects, enables the description of general pseudospin control including pseudospin flipping. Bare models are then reduced to an effective valence band theory encapsulating the low-energy SVO physics including linear response of pseudospins to external fields. Based on the theory, it discusses Rashba and Zeeman type effects in electric and magnetic fields, respectively, of arbitrary orientations. Two paradigm configurations of static external fields are identified for pseudospin control, with one involving only vertical fields and the other in-plane fields. For an example of applications, spin-valley-orbital based quantum computing is proposed, with qubits formed of quantum dot-confined holes. The theory is applied to address the challenge in all-electrical, simultaneous quantum control of spin, valley, and atomic orbital indices for qubit manipulation, and demonstrate the viability of such control with an additional dynamic, in-plane electric field in both configurations. An optimized qubit manipulation time ~ O(*ns*) is given.

This paper is organized as follows. To prepare for the whole discussion of the work, **Sec. II** introduces elastic valley-flip scattering. **Sec. III** presents the symmetry perspective of SVO physics in external fields, and demonstrates the two configurations of interest for pseudospin control. **Sec. IV** presents bare models and the main result - effective theory of field-modulated SVO physics, with a discussion of Rashba and Zeeman type field effects. **Sec. V** presents the SVO-based quantum computing - qubit states, and qubit manipulation via external field modulation. **Sec. VI** concludes the study. **Appendix A** summarizes a few important matrix elements used in this work. **Appendix B** provides a supplement of certain mathematical details for bare models. **Appendix C** summarizes the main theoretical tool of this work − Schrieffer-Wolff reduction, and applies it to the derivation of effective theory, as well as systems of dynamic electric field-driven qubits. Expressions of coupling parameters in the theory in terms of bare ones are derived. **Appendix D** presents a discussion of elastic scattering, including both valley-conserving and valley-flipping ones that enter bare models.

## II. ELASTIC VALLEY-FLIP SCATTERING

For complete pseudospin control, one must be able to "rotate" a pseudospin arbitrarily, in the two-state space expanded by $\{|K>, |K'>\}$. This includes the pseudospin flip $|K> \leftrightarrow |K'>$ as an important type of manipulation. As such flip consists partially of reversing the valley index, the existence of a mechanism to couple opposite valleys, or flip valley, is a necessary condition for complete pseudospin control.

Elastic carrier scattering can change the wave vector and compensate for the difference between K and K', providing valley-flip coupling. Such scattering occurs spontaneously at impurities or, in a more controlled fashion, at boundaries in quantum structures. We denote $U_{elastic}$ as the corresponding scattering potential energy.

For quantum structures, we focus specifically on the armchair nanoribbon-based quantum wires (QWs) and quantum dots (QDs) with confinement potentials shown in **Figure 2**. In these structures, since the wave vectors at K and K' are normal to the armchair edge, the edge scattering can effectively provide the wave vector difference needed for valley flip. In the case of QDs, the scattering can be optimized using a triangular QD with all armchair edges [61]. For a similar purpose, a sharp confining boundary is preferred over a graded one. In this work, however, we do not attempt to maximize valley-flip scattering. Instead, we focus on structures with an intermediate coupling, for example, a rectangular QD defined by sharp armchair edges and graded zigzag edges, with a quadratic confining potential profile in association with graded edges as shown in **Figure 2**. Such structures allow for

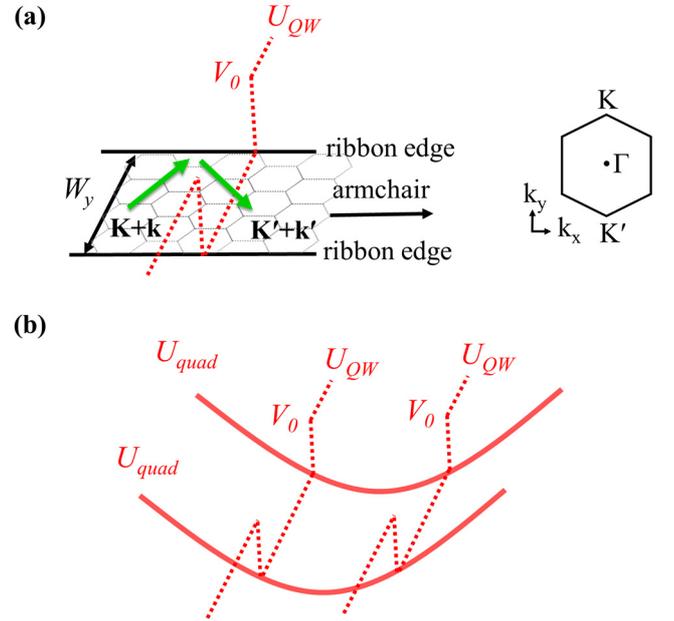

**Figure 2** (a) Illustration of an armchair nanoribbon and the 1$^{st}$ Brillouin zone of underlying bulk lattice. The confinement potential $U_{QW}$ is taken to be piecewise constant. A K-electron with wave vector "**K+k**" is scattered into a K'-electron with wave vector "**K'+k'**". (b) Illustration of the QD potential profile, which consists of a harmonic potential ($U_{quad}$) in the armchair direction and $U_{QW}$ in the zigzag direction.

an analytical treatment as well as possible experimental realization, and the corresponding study should be sufficiently informative for assessing general quantum structures.

In general, all-gate patterning technique may be applied to electrostatically define nanoribbons or QDs, depending on the availability of advanced lithography facilities with sharp, lateral pattern defining capacity.[48-52] In a somewhat varied approach, the QD may be gate-patterned in an already grown armchair nanoribbon, with ribbon edges serving as boundaries of the QD on two sides. This approach would require a passivation of the surface states [62] on armchair edges. In yet another alternative, the QD may be fabricated in a lateral TMDC-based heterostructure, where the valence band offset between materials serves to confine a hole. [63,64]. Overall, in general quantum structures, $U_{elastic}$ can include both electric gate-induced confinement potential and valence band offset.

In this presentation, $U_{elastic}$ is taken to be nonmagnetic and an even function under the reflection $z \rightarrow -z$. Generalization of the theory to arbitrary $U_{elastic}$ is possible at the cost of increased presentation complexity. Specifically, we consider

(i) $$U_{elastic} = \sum_{R_i} v_{impurity}(\vec{r} - \vec{R}_i), \quad (2)$$

where $R_i$ denotes impurity position, in the case of a bulk with a random, dilute distribution of identical impurities;

(ii) $$\begin{aligned} U_{elastic} &= U_{QW}(y) \\ &= V_0 \left[ \theta(-W_y/2 - y) + \theta(y - W_y/2) \right], \end{aligned} \quad (3)$$

in the case of a nanoribbon, where $V_0$ is the barrier height and $W_y$ is the y-dimension; or

(iii) $$\begin{aligned} U_{elastic} &= U_{QD}(x,y) \\ &= U_{quad}(x) + U_{QW}(y), \\ U_{quad}(x) &= (1/2) m^* \omega_x^2 x^2, \end{aligned} \quad (4)$$

in the case of a QD. The harmonic potential energy $U_{quad}(x)$ provides the x-confinement and gives a corresponding x-dimension $W_x \sim (\hbar/m^* \omega_x)^{1/2}$ ($m^*$ = hole effective mass = $O(m_e)$; $m_e$ = electron mass in vacuum; $\omega_x$ = frequency parameter for the harmonic potential). In practical quantum structures, $U_{elastic}$ in (ii) and (iii) is defined basically with a unit-cell scale resolution, meaning that $U_{elastic}$ actually varies insignificantly in a unit cell.

## III. SYMMETRY PERSPECTIVE

For a SVO pseudospin, with more DoFs than just valley involved, the elastic scattering mechanism alone is insufficient to flip such pseudospin. In the case of nanoribbons, due to the insufficiency, energy subbands are always valley-polarized in spite of the ribbon edge scattering.[62] We provide an analysis below for such valley rigidity, show that it has a symmetry origin, and demonstrate configurations of external fields that can successfully break the symmetry and lift the rigidity, effecting a pseudospin-flip coupling for pseudospin control.

**Vertical configuration**

This configuration consists primarily of a static, vertical electric field $\varepsilon_z$. The following explains the role of $\varepsilon_z$ in symmetry breaking.

We use an armchair nanoribbon for the discussion. When free of external fields, it has the symmetry of time-reversal ($T$), and mirror reflection with respect to the layer plane ($M_z$) as well as the center axis ($M_y$). When $\varepsilon_z \neq 0$, $M_y$ and $T$ are preserved but $M_z$ is broken.

For $\varepsilon_z = 0$, energy eigenstates are valley-degenerate and denoted as $|K, k_x, n\rangle$ and $|K', k_x, n\rangle$ ($n$ = subband index, and $k_x$ = wave vector in the x-direction). In the general case where $\varepsilon_z$ may be finite, it can be shown that the common eigenstates of both energy and $M_y$ can be written in the following forms

$$|\pm, k_x\rangle_y = \sum_m C_m(\varepsilon_z) \left[ |K, k_x, m\rangle \pm i |K', k_x, m\rangle \right], \quad (5)$$

with

$$M_y |+, k_x\rangle_y = |+, k_x\rangle_y, \quad M_y |-, k_x\rangle_y = -|-, k_x\rangle_y. \quad (6)$$

Eqn. (5) expresses a possible occurrence of mixing between subbands when $\varepsilon_z \neq 0$. For $\varepsilon_z = 0$, the mixing vanishes, and it reduces to the simple result where $C_n = 1$ and $C_{m \neq n} = 0$ for a certain subband of index "$n$", for example. Above, a subscript "$y$" is attached to the state to indicate that the pseudospin is "polarized in the y-direction", as implied by Eqn. (6).

Under the $M_z$ operation, the above eigenstates transform into each other, with

$$M_z |+, k_x\rangle_y = |-, k_x\rangle_y, \quad M_z |-, k_x\rangle_y = |+, k_x\rangle_y. \quad (7)$$

Eqn. (7) implies the following. For $\varepsilon_z = 0$, with $M_z$ a symmetry of the system, the equation constrains $|+, k_x\rangle_y$ and $|-, k_x\rangle_y$ or, the corresponding basis states - $|K, k_x, n\rangle$ and $|K', k_x, n\rangle$ for example - to be degenerate. For $\varepsilon_z \neq 0$, $M_z$ becomes broken, invalidating the constraint. Effectively, it implies a possible energy splitting between $|+, k_x\rangle_y$ and $|-, k_x\rangle_y$ or, equivalently, a coupling between the basis states $|K, k_x, n\rangle$ and $|K', k_x, n\rangle$. Such coupling can then be exploited for the flip manipulation $|K\rangle \leftrightarrow |K'\rangle$. We denote the coupling as $H_{\varepsilon_z}^{(Rashba)}$, and discuss its nature next.

Under the $T$ operation, we have

$$T |+, k_x\rangle_y = i |-, -k_x\rangle_y, \quad T |-, k_x\rangle_y = -i |+, -k_x\rangle_y, \quad (8)$$

which constrains $\{|+, k_x\rangle_y, |-, -k_x\rangle_y\}$ to be degenerate. When combined with the possible $|+, k_x\rangle_y - |-, k_x\rangle_y$ splitting, it implies the existence of a Rashba-type energy splitting between "+" and "-" bands. A numerical tight-binding calculation verifies this expectation, as shown in **Figure 3 (a)**, which presents Rashba-split subbands. From such splitting, $H_{\varepsilon_z}^{(Rashba)}$ can be deduced with a simple perturbation-theoretical argument for two-state systems, which gives the following Rashba-type form

$$H^{(Rashba)}_{\varepsilon_z} \propto \varepsilon_z k_x \quad (9)$$

in the leading order when $k_x \sim 0$.

The requirement of fields varies in the case of a QD, as shown in the top graph of **Figure 3(a)**. Two additional fields are introduced. Firstly, in a QD it has the vanishing expectation value $<H^{(Rashba)}_{\varepsilon_z}>_{QD} \propto <k_x>_{QD} = 0$ for the coupling. To get around the issue, an ac electric field $\varepsilon_{ac}$ in the $x$-direction, with frequency $\omega_{ac}$, is introduced into the configuration. Secondly, for the ac field to work effectively, the carrier must be in resonance with $\varepsilon_{ac}$. Therefore, a vertical magnetic field $B_z$ is further included to Zeeman-split $|K>$ and $|K'>$, with the corresponding Larmor frequency $\omega_L^\perp = $ (Zeeman energy)$/\hbar \approx \omega_{ac}$ satisfying the resonance condition. **Sec. V** provides more details, when demonstrating a complete pseudospin control in the QD case.

**In-plane configuration**

This configuration consists primarily of a static, in-plane magnetic field. We again use the nanoribbon as an example, and take the magnetic field in the $x$-direction ($B_x$). With the field, the spin Zeeman interaction $\propto s_x B_x$ is introduced into the system ($\vec{s}$ = Pauli spin operator), which can flip a spin and hence assist the $|K>$-$|K'>$ coupling. Other magnetic effects, e.g., the Landau orbital quantization, cannot directly induce pseudospin flip and, hence, would only produce higher-order corrections.

In the presence of Zeeman interaction, the composite $M_zM_y$ and $TM_y$ are both symmetry elements of the system. For example, when applying to a spin, $M_zM_y \sim s_zs_y$ and thus commutes with $s_xB_x$.

The common eigenstates of energy and $M_zM_y$ are given by

$$|\pm, k_x\rangle_x = \sum_n C_n(B_x)\big[|K, k_x, n\rangle \pm |K', k_x, n\rangle\big], \quad (10)$$

with

$$M_zM_y |+, k_x>_x = -i |+, k_x>_x, \; M_zM_y |-, k_x>_x = i |-, k_x>_x. \quad (11)$$

The subscript "$x$" above indicates that the pseudospin is "polarized in the $x$-direction".

Under $TM_y$, we have

$$TM_y |+, k_x>_x = i |+, -k_x>_x, \; TM_y|-, k_x>_x = i |-, -k_x>_x. \quad (12)$$

This constrains $\{|+, k_x>_x, |+, -k_x>_x\}$ as well as $\{|-, k_x>_x, |-, -k_x>_x\}$ to be pairs of degenerate states. Therefore, the subbands show a $B_x$-induced Zeeman-type splitting, denoted as "$\hbar\omega_L^{//}$", between the "+" and "-" states. This expectation is confirmed by a numerical tight-binding calculation, as shown in **Figure 3 (b)**, which presents Zeeman-split subbands. With a perturbation-theoretical argument, it points to the existence of $B_x$-induced $|K>$-$|K'>$ coupling.

In the in-plane configuration, electric pseudospin control can be achieved by creating an electric coupling between "+" and "-" bands with, for example, the VOI derived from a static electric field $\varepsilon_y$. Based on Eqn. (1), $H_{VOI} \propto \tau\varepsilon_y k_x$ and, hence,

$$<+, k_x| H_{VOI} |-, k_x>_x \propto \varepsilon_y k_x, \quad (13)$$

giving a coupling between the "+" and "-" states.

In the case of a QD, the coupling vanishes because $<k_x>_{QD} = 0$. One can again solve the issue by introducing into the configuration an ac electric field $\varepsilon_{ac}$ in the $x$-direction, with the frequency $\omega_{ac}$ satisfying the resonance condition $\omega_{ac} \approx \omega_L^{//}$, as discussed in **Sec. V**. The overall field configuration is shown in the top graph of **Figure 3(b)**.

The above symmetry-based analysis not only yields useful configurations for pseudospin control, it also sets up a constraint on the construction of effective theory - the theory should incorporate correct symmetry and reproduce the same symmetry-breaking phenomena demonstrated above, as we proceed to the next section and present the theory.

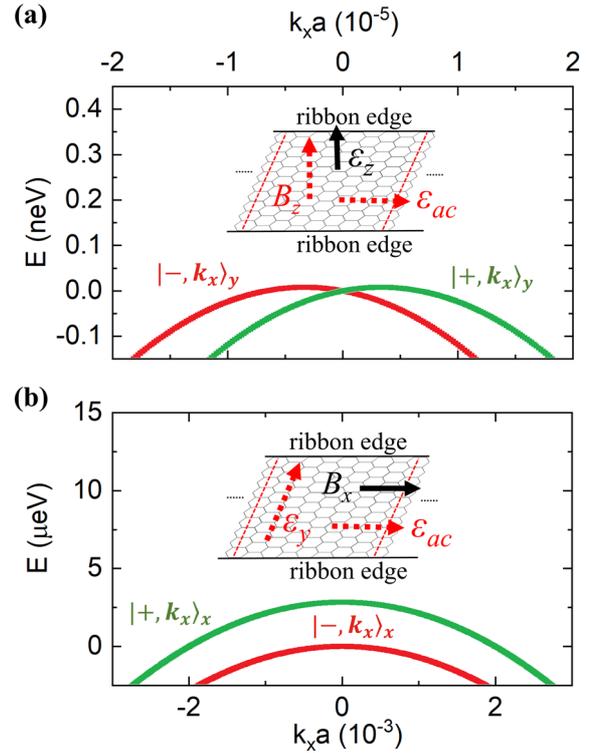

**Figure 3 (a)** Subband structure with Rashba type energy splitting, in a WSe$_2$ armchair nanoribbon in the vertical-field configuration, with $\varepsilon_z = 10\ mV/a$ (solid black arrow) and $W_y = 9\ a$. **(b)** Subband structure with Zeeman type energy splitting, in a WSe$_2$ armchair nanoribbon in the in-plane-field configuration, with $B_x = 1\ T$ (solid black arrow) and $W_y = 9\ a$. The tight-binding model parameters here are adopted from References 39 and 42. Red thin dashed lines portray additional confinement, besides that provided by ribbon edges, for a QD. Red dashed arrows denote additional fields required for pseudospin control in the QD structure, in the two configurations.

**IV. THE EFFECTIVE THEORY**

Valence (conduction) band edge states in a TMDC crystal are primarily composed of $d_{\pm 2}$ ($d_0$) orbitals of the metal atom with the symmetry of even parity under $M_z$. However, as the pseudospin flip $|K\rangle \leftrightarrow |K'\rangle$ requires multiple quantum index mixing, it generally involves plural intermediate states both near and distant from the band gap which derive from, besides $d_0$ and $d_{\pm 2}$, also $d_{\pm 1}$ with odd parity under $M_z$. **Figure 4** summarizes the TMDC band structure, with a tabulation of band edge states at Dirac points both near and away from the gap. It describes the symmetry of their wave functions in terms of the quasi-atomic orbital notations, e.g., $D_0$, $P_0$, $D_{\pm 2}$ and $D_{\pm 1}$, with corresponding wave functions $\Psi_{D_0} \sim z^2$, $\Psi_{P_0} \sim z$, $\Psi_{D_{\pm 2}} \sim (x \pm iy)^2$ and $\Psi_{D_{\pm 1}} \sim (x \pm iy)z$. (Lower-cased letters "p" and "d" are reserved for true atomic orbitals.) These notations of ours correspond to the standard group irreducible representations (IR) $A_1$, $A_2$, $E_{1\pm}$ and $E_{2\mp}$, respectively, of $C_{3h}$ – the 2D hexagonal symmetry group, and are introduced here to describe states and specify in particular their transformation properties under $C_{3h}$ symmetry operations. For example, $|VBM, K\rangle$ and $|VBM, K'\rangle$ have IR indices $D_2$ (or $E_{1+}$) and $D_{-2}$ (or $E_{1-}$), respectively, in our (standard) notations. The figure also presents the primary constituent atomic orbital of each state, e.g., $d_0$, $d_{\pm 2}$ and etc. As it shows, they are closely correlated with the corresponding IR indices $D_0$, $D_{\pm 2}$ and etc., justifying the quasi-atomic orbital notations introduced by us. However, the correlation breaks down when atomic p-orbitals of chalcogen (X) are primary, due to the following reason. In our convention, the metal ion (M) is taken to be the center about which one performs a symmetry operation. Therefore, in the case of p-orbitals the correlation would hold if they belong to M but would not if they belong to X. Overall, the IR index specifies the wave function symmetry of a state with respect to the metal ion.

cased letters "P" and "D" denote corresponding irreducible representation indices of states. "c" denotes the bottom conduction band, "c+1" the next conduction band, and etc. Here, d-orbitals come from metal M while p-orbitals from chalcogen X.

Nontrivial elements of $C_{3h}$ consist of $C_3$ and $M_z$, with $C_3$ the three-fold rotation and $M_z$ the mirror reflection with respect to x-y plane. **Table 1** tabulates transformation properties of various states under $C_3$ and $M_z$ as well as the correspondence between our and standard group-theoretical notations.

| $C_{3h}$ irreducible representation (standard notation) | $A_1$ | $A_2$ | $E_{1\pm}$ | $E_{2\pm}$ |
|---|---|---|---|---|
| Our notation | $D_0$ | $P_0$ | $D_{\pm 2}$ | $D_{\mp 1}$ |
| State symmetry | $\sim z^2$ | $\sim z$ | $\sim (x \pm iy)^2$ | $\sim (x \mp iy)z$ |
| Symmetry operation $C_3$ | 1 | 1 | $\omega^\pm$ | $\omega^\pm$ |
| Symmetry operation $M_z$ | 1 | -1 | 1 | -1 |

**Table 1** Summary of the transformation of various states and the correspondence between our and the standard group-theoretical notations. $\omega^\pm = e^{\pm i2\pi/3}$.

Under $C_3$ and $M_z$, the states are transformed as follows:

(1) $C_3 \Psi_{D_0(P_0)} = \Psi_{D_0(P_0)}$, $C_3 \Psi_{D_{\mp 1(\pm 2)}} = \omega^\pm \Psi_{D_{\mp 1(\pm 2)}}$, where $\omega^\pm = e^{\pm i2\pi/3}$;

(2) $M_z \Psi_{D_0} = \Psi_{D_0}$, $M_z \Psi_{P_0} = -\Psi_{P_0}$, $M_z \Psi_{D_{\pm 2}} = \Psi_{D_{\pm 2}}$, $M_z \Psi_{D_{\pm 1}} = -\Psi_{D_{\pm 1}}$.

A knowledge of the state symmetry and transformation properties is useful when calculating matrix elements between the states. Some key matrix elements used in this work are given in **Appendix A**.

The effective theory of SVO physics is presented below, in the context of complicated TMDC band structure summarized in **Figure 4**. It accounts for field effects in electric $\boldsymbol{\varepsilon}$ ($\boldsymbol{\varepsilon} = (\vec{\varepsilon}_{//}, \varepsilon_z)$, $\vec{\varepsilon}_{//} = (\varepsilon_x, \varepsilon_y)$) and magnetic **B** (**B** = ($\vec{B}_{//}$, $B_z$), $\vec{B}_{//} = (B_x, B_y)$) fields, in the linear regime, and provides a theoretical framework for field modulation-based studies and applications in SVO physics.

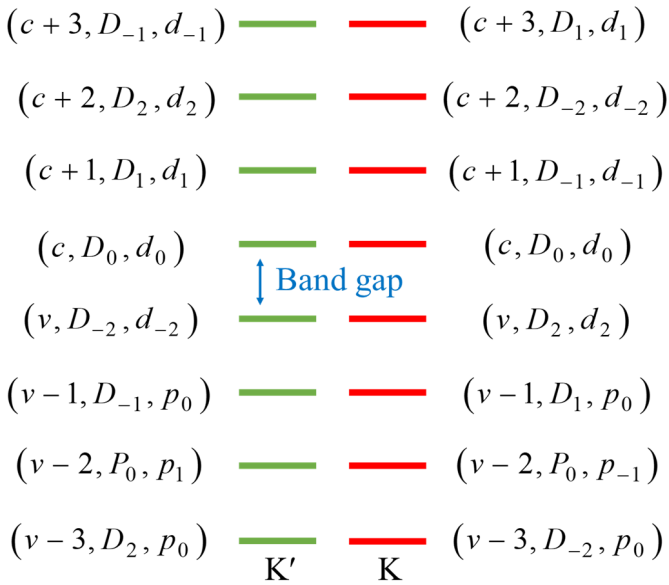

**Figure 4** Irreducible representations and atomic orbital characters of valence and conduction band states at K and K', based on TMDC band structure calculations, e.g., Reference 47. Lower-cased letters "p" and "d" denote primary constituent atomic orbitals of states while upper-

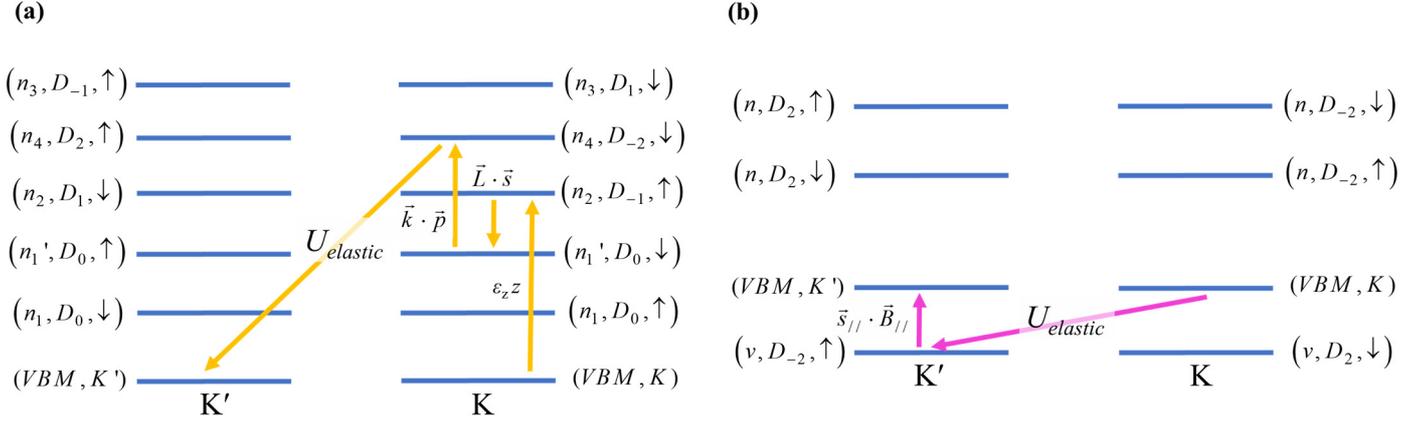

**Figure 5** Examples of leading-order quantum paths for SVO pseudospin flipping ($|K\rangle \to |K'\rangle$), in the presence of **(a)** $\varepsilon_z$ and **(b)** $\vec{B}_{//}$. **(a)** shows a four-step, **Class A** quantum path (brown arrows) involving parity flip $(VBM, K) \to (n_2, D_{-1}, \uparrow, K)$, SOI coupling ($\propto \vec{L} \cdot \vec{s}$) $(n_2, D_{-1}, \uparrow, K) \to (n_1', D_0, \downarrow, K)$, $\vec{k} \cdot \vec{p}$ coupling [65] $(n_1', D_0, \downarrow, K) \to (n_4, D_{-2}, \downarrow, K)$, and elastic scattering $(n_4, D_{-2}, \downarrow, K) \to (VBM, K')$. **(b)** shows a two-step, **Class B** path (purple arrows), which consists of elastic scattering $(VBM, K) \to (v, D_{-2}, \uparrow, K')$ and spin flip $(v, D_{-2}, \uparrow, K') \to (VBM, K')$. Here, the notations $|VBM, K\rangle (= (v, D_2, \uparrow, K))$ and $|VBM, K'\rangle (= (v, D_{-2}, \downarrow, K'))$ are used in place of $|K\rangle$ and $|K'\rangle$, respectively, to explicitly indicate their locations at the valence band maximum. $(n_1, D_0, \uparrow)$ and etc. denote intermediate states. "$n_1$" and etc. are band indices. Specifically, "$v$" = valence band. $\hbar \vec{L}$ = angular momentum operator, and $\vec{s}$ = Pauli spin operator ($\vec{s}_{//} = (s_x, s_y)$)

The theory is intended to cover both intra- and inter-valley electron dynamics, with the inter-valley part describing the $|K\rangle$-$|K'\rangle$ coupling. **Figure 5** illustrates some of leading-order contributions to the coupling, by showing corresponding quantum paths and intermediate states involved. Basically, for a SVO pseudospin to flip, spin, valley, and wave function symmetry (or IR) indices of the pseudospin must all switch. As the figure shows, the IR index can be flipped by either $U_{elastic}$-induced scattering or other couplings. Depending on whether the IR index is conserved or not during $U_{elastic}$-induced scattering, the paths are classified into **Class A** and **Class B** as demonstrated in the figure: **Class A** of "IR-diagonal" nature and **Class B** of "IR-flipped" nature.

**IV-1** presents the theory and then discusses important field effects based on the theory. **IV-2** describes bare models. **IV-3** provides expressions of effective coupling parameters.

### IV-1. Theory and field effects

The theory describes the quantum mechanics of near-band-edge valence band states, in the pseudospin state space expanded by $\{|K\rangle, |K'\rangle\}$. A general state in the space is expressed as $|\Psi\rangle = F_K |K\rangle + F_{K'} |K'\rangle$, where $F_K$ and $F_{K'}$ are envelope functions. They are governed by the following wave equation

$$H_{eff} \begin{pmatrix} F_K \\ F_{K'} \end{pmatrix} = E \begin{pmatrix} F_K \\ F_{K'} \end{pmatrix}, \quad (14)$$

where $H_{eff}$ is the Hamiltonian. We divide $H_{eff}$ into diagonal (pseudospin-conserving) and off-diagonal (pseudospin-flipping) parts, i.e.,

$$H_{eff} = H_{eff}^{(diag)} + H_{eff}^{(off-diag)}. \quad (15)$$

Each part is presented and discussed below. A number of coupling parameters are present in $H_{eff}$ and reflect the existence of rich physics in the pseudospin space. Overall, five primary ones, $\{g_{eff}^{\perp}, g_{eff}^{(2)//}, g_{eff}^{(3)//}, R_{VOI,eff}, R_{SOI,eff}^{(4)\perp}\}$, and two secondary ones, $\{g_{eff}^{(3,corr)//}, R_{SOI,eff}^{(4,corr)\perp}\}$, characterize $H_{eff}$ as well as the linear response of a pseudospin to external fields, with secondary parameter-dependent Hamiltonian terms taken to be "corrections", as they are dominated by corresponding primary ones (see **Appendix D**). Exact role of each parameter will become clear below. Expressions of these parameters are presented below in **IV-3** and **Appendix C**.

The diagonal part governs the intra-valley dynamics and is given by

$$H_{eff}^{(diag)} = \left( \vec{\Pi}^2 / 2m^* + e\vec{\varepsilon}_{//} \cdot \vec{r} + U_{elastic} \right) \mathbf{1} + H_{VOI}^{(val)} + \frac{1}{2} E_{Z,eff}^{\perp} v_z. \quad (16)$$

($\vec{\Pi} = \vec{p} + e\vec{A}$; $\vec{A}$ = vector potential due to $B_z$, and $(v_x, v_y, v_z)$ = Pauli "pseudospin" operator in the pseudospin state space. Specifically, $v_z = \begin{pmatrix} 1 & 0 \\ 0 & -1 \end{pmatrix}$ in the basis $\{|K\rangle, |K'\rangle\}$, and etc.) The first term describes the orbital part of dynamics, in external fields $\vec{\varepsilon}_{//}$ and $B_z$, and the potential $U_{elastic}$. The remaining terms predict two important field effects on the pseudospin part.

**In-plane Rashba effect due to $\vec{\varepsilon}_{//}$**

$H_{VOI}^{(val)}$ is the valley-orbit interaction given by

$$H_{VOI}^{(val)} = R_{VOI,eff}\left(\vec{\Pi} \times \vec{\varepsilon}_{//}\right)_z v_z. \quad (17)$$

Let $\varepsilon = \varepsilon_y \hat{y}$, and $\mathbf{B} = 0$. Then $H_{VOI}^{(val)}$ leads to a Rashba energy term

$$<\tau, k_x, k_y | H_{VOI}^{(val)} | \tau, k_x, k_y>$$
$$= \tau \alpha^{//} k_x,$$
$$\alpha^{//} = \hbar R_{VOI,eff} \varepsilon_y \quad (18)$$

for bulk states ($\tau = 1$ (or K) or -1 (or K')), giving an energy splitting "$2\alpha^{//} k_x$" between the states $|K, k_x, k_y>$ and $|K', k_x, k_y>$. Here, $\alpha^{//}$ is the corresponding Rashba effect constant. With a similar argument, this effect exists in the nanoribbon case between subband states of opposite valleys, e.g., $|K, k_x, n>$ and $|K', k_x, n>$ ($n$ = subband index) in both armchair and zigzag nanoribbons.

As the effect arises out of the VOI, $R_{VOI,eff}$ is the only relevant coupling parameter in the effect.

**Vertical Zeeman effect due to $B_z$**

$E_{Z,eff}^\perp = g_{eff}^\perp \mu_B B_z$ in the last term of $H_{eff}^{(diag)}$. For $\varepsilon = 0$ and $\mathbf{B} = B_z \hat{z}$, it results in a Zeeman type splitting "$E_{Z,eff}^\perp$" between $|K, k_x, k_y>$ and $|K', k_x, k_y>$ in the bulk case as well as between corresponding subband states in both armchair and zigzag nanoribbons, with $g_{eff}^\perp$ the corresponding g-factor in this effect. $g_{eff}^\perp$ consists of two parts, namely, "$g_e$" and "$g_{valley-orbital}$". $g_e$ derives from the existence of $\vec{\mu}_s$, giving $g_e = 2$, while $g_{valley-orbital}$ from that of $\vec{\mu}_\tau$ and $\vec{\mu}_L$. For the expression of $g_{valley-orbital}$, see **IV-3**.

Note that $g_{eff}^\perp$ is the only relevant coupling parameter in the present effect.

The off-diagonal part of $H_{eff}$ is given by

$$H_{eff}^{(off-diag)}$$
$$= -iv_+\left\{\frac{1}{\Delta} R_{SOI,eff}^{(4)\perp}\varepsilon_z \left\{\left\{\Pi_+, U_{elastic}^{(IR-diag)} e^{-2iKy}\right\}_+\right.\right.$$
$$\left.+ \left\{\Pi_-, R_{SOI,eff}^{(4,corr)\perp}\left(U_{elastic}^{(IR-flip)}\right)_- e^{-2iKy}\right\}_+\right\}$$
$$+ \frac{1}{\Delta}\mu_B B_-\left\{ig_{eff}^{(2)//}\left(U_{elastic}^{(IR-flip)}\right)_- e^{-2iKy}\right.$$
$$+ \frac{a}{\hbar} g_{eff}^{(3)//}\left\{\left[\Pi_-, U_{elastic}^{(IR-diag)} e^{-2iKy}\right]\right.$$
$$\left.\left.\left.+ \left[\Pi_+, g_{eff}^{(3,corr)//}\left(U_{elastic}^{(IR-flip)}\right)_+ e^{-2iKy}\right]\right\}\right\}\right\}$$
$$+ h.c. \quad (19)$$

($B_\pm = B_x \pm iB_y$, $\Pi_\pm = \Pi_x \pm i\Pi_y$, $\overline{\Delta}$ = typical energy gap, $v_+ = \frac{1}{2}(v_x + iv_y) = \begin{pmatrix} 0 & 1 \\ 0 & 0 \end{pmatrix}$ and etc.) $H_{eff}^{(off-diag)}$ describes the $|K>$-$|K'>$ coupling and involves quite a few coupling parameters, namely, $\{g_{eff}^{(2)//}, g_{eff}^{(3,corr)//}, g_{eff}^{(3)//}, R_{SOI,eff}^{(4)\perp}, R_{SOI,eff}^{(4,corr)\perp}\}$. The superscript number of a parameter denotes the perturbation-theoretical order of quantum paths involved in the corresponding Hamiltonian term. Several valley-flipping potential energy functions appear in $H_{eff}^{(off-diag)}$. They are derived from $U_{elastic}$, with the superscript "IR-diag" ("IR-flip") indicating that the underlying valley-flip scattering conserves (changes) the irreducible representation index of electron state. Such derived functions are obtained in **Appendix D** and given below in a few cases of interest.

(i) In the case of a bulk with random, dilute distribution of identical, short-range impurities on M sites,

$$U_{elastic}^{(IR-diag)}(\vec{r}) = \sum_{R_i} v_{impurity}(0)\delta_{\vec{R}(\vec{r}),\vec{R}_i},$$

$$\vec{U}_{elastic}^{(IR-flip)}(\vec{r}) = U_{elastic}^{(IR-diag)}\hat{x},$$

$$\left(U_{elastic}^{(IR-flip)}\right)_\pm = U_{elastic}^{(IR-diag)}. \quad (20)$$

$\vec{R}(\vec{r})$ is the lattice site nearest to $\vec{r}$. Basically, we do not distinguish between IR- flipped and conserved potentials in this case.

(ii) In the case of quantum structures,

$$U_{elastic}^{(IR-diag)} \approx U_{elastic},$$

$$\vec{U}_{elastic}^{(IR-flip)} \approx a\vec{\nabla}U_{elastic},$$

$$\left(U_{elastic}^{(IR-flip)}\right)_\pm \approx a\left(\partial_x \pm i\partial_y\right)U_{elastic} \quad (21)$$

("$a$" = lattice constant).

The lengthy expression of $H_{eff}^{(off-diag)}$ has a number of important mathematical features, which can be interpreted from physics point of view, as follows. These features are mostly closely connected with the pseudospin-flipping nature of $H_{eff}^{(off-diag)}$.

(i) Explicit vertical $\varepsilon_z$- and in-plane $\vec{B}_{//}$- dependences: This feature agrees with the result in **III** of field configurations derived with symmetry-based analysis for pseudospin flipping manipulation.

(ii) Presence of the common factor "$e^{-2iKy}$" throughout

the expression:
This comes from the need to compensate for wave vector difference between |K> and |K'> in the flipping.

(iii) Presence of valley-flipping potential energy functions throughout the expression:

In particular, when $\vec{U}_{elastic}^{(IR-flip)}(\vec{r}) = U_{elastic}^{(IR-diag)} = 0$, $H_{eff}^{(off-diag)} = 0$, implying the absence of any pseudospin flipping, as we would expect, for example, in the trivial case of a defect-free bulk.

(iv) Presence of the momentum operator $\vec{\Pi}$ up to the first order:

Being a low-energy theory, $H_{eff}$ is primarily valid in the vicinity of Dirac points. As will become clear later, its derivation based on the Schrieffer-Wolff reduction involves a perturbation - the "$\vec{k} \cdot \vec{p}$" term up to the first order ($\vec{k}$ = wave vector relative to the nearest Dirac point). When making the effective-mass approximation with the substitution $\hbar \vec{k} \to \vec{\Pi}$ in the derivation, it results in the presence of $\vec{\Pi}$ in $H_{eff}^{(off-diag)}$ also up to the same order.

(v) Separation of $U_{elastic}$ into "IR-diag" and "IR-flip" components:

Quantum path **Classes A** and **B** have different structures since they involve distinct scattering, namely, IR -conserved and -flipped ones, respectively. This results in a corresponding difference in the functional forms of derived Hamiltonian terms, as is manifested in, for example, $U_{elastic}^{(IR-diag)}$ - and $\left(U_{elastic}^{(IR-flip)}\right)_{-}$ - dependent terms of the third-order perturbation-theoretical order in $H_{eff}^{(off-diag)}$, which entangle with different operators, e.g., $\Pi_{-}$ and $\Pi_{+}$, respectively. This explains why the separation of $U_{elastic}$ into "IR-diag" and "IR-flip" components as well as a corresponding classification of quantum paths into **Classes A** and **B** naturally enters the formulation.

(vi) Presence of anti-commutators "$\{....\}_{+}$" in the $\varepsilon_z$ - dependent term and commutators "$[....]$" in the $\vec{B}_{//}$ - dependent term:

This difference in algebra leads to distinct functional forms of the Rashba and Zeeman effects in association with $\varepsilon_z$ and $\vec{B}_{//}$. While a detailed discussion of the effects will be presented below, here we briefly explain the correlation between the algebra and the effects. Let $k_x$ be a quantum index of the electron. With anti-commutators "$\{....\}_{+}$" in the $\varepsilon_z$-dependent term, it is expected that $\{k_x, ....\}_{+} \propto k_x$, giving a linear-in-$k_x$ dependence in the $\varepsilon_z$-induced energy for $k_x \sim 0$, which complies with the functional form – being odd in $k_x$ of Rashba splitting obtained in **III**. While with commutators "$[....]$" in the $\vec{B}_{//}$-dependent term, it is expected that $[k_x, ....] \sim 0$, hence forbidding any linear-in-$k_x$ dependence in the $\vec{B}_{//}$-induced energy for $k_x \sim 0$, which complies with the functional form – being even in $k_x$ of Zeeman splitting obtained in **III**.

Below we consider the case of an armchair nanoribbon in the $x$-direction, for which Eqn. (21) shows the presence of confinement-induced $\vec{U}_{elastic}^{(IR-flip)}(\vec{r})$ and $U_{elastic}^{(IR-diag)}$. Therefore, nontrivial consequences rising from pseudospin flipping are expected. Specifically, Rashba and Zeeman effects due to $\varepsilon_z$- and $\vec{B}_{//}$, respectively, will be demonstrated, with detailed mathematical expressions provided and shown to agree with the result derived in **III**.

**Vertical Rashba effect due to $\varepsilon_z$**

For $\boldsymbol{\varepsilon} = \varepsilon_z \hat{z}$ and **B** = 0, a coupling between subband states |K, $k_x$, n> and |K', $k_x$, n> exists and is given by

$$< K, k_x \sim 0, n | H_{eff}^{(off-diag)} | K', k_x \sim 0, n > = i\alpha^{\perp} k_x. \quad (22)$$

Above,

$$\alpha^{\perp} \approx R_{SOI,eff}^{(4)\perp} \varepsilon_z \left(\frac{\hbar^2 k_{y,n}^2}{2m^*}\right) \left\{-\frac{4}{\Delta} \frac{\hbar^2}{\sqrt{2m^*V_0}} \frac{\cos(KW_y)}{W_y} + \frac{8\hbar}{\overline{\Delta}} \left(R_{SOI,eff}^{(4,corr)\perp}\right) \frac{a\sin(KW_y)}{W_y}\right\} \quad (23)$$

is the leading-order Rashba effect constant in the hard-wall limit where $V_0 \gg \frac{\hbar^2 k_{y,n}^2}{2m^*}$ ($k_{y,n} = (n+1)\pi/W$). Thus, in the subspace expanded by $\{|\tau, k_x, n>$'s, $\tau$ = K, K'$\}$,

$$H_{eff}^{(off-diag)} = -\alpha^{\perp} k_x v_y, \quad (24)$$

where $v_y$ is the Pauli operator in the subspace. Due to the coupling, energy eigenstates in the subspace are given by $|+, k_x, n>_y$ and $|-, k_x, n>_y$ with a Rashba-split subband structure, in agreement with the result shown in **Figure 3(a)**. Our theory yields a linear $k_x$ energy splitting "$2\alpha^{\perp} k_x$" for states near $k_x$ = 0. In addition, Eqn. (23) predicts an oscillatory and decaying behavior in the energy splitting when increasing $W_y$. Such prediction is numerically confirmed by the same tight-binding calculation used to obtain **Figure 3 (a)**.

In Eqn. (22), subband state wave functions in the hard-wall limit are given by

$$|\tau, k_x, n> = (1/W_x)^{1/2} e^{ik_x x} Y_n(y) | VBM, \tau>,$$
$$Y_n(y) = (2/W_y)^{1/2} \begin{cases} \cos(k_{y,n} y), n = 2n', \\ \sin(k_{y,n} y), n = 2n'+1, \end{cases} \quad (25)$$

which will be used again below.

Eqn. (23) indicates $\{R_{SOI,eff}^{(4)\perp}, R_{SOI,eff}^{(4,corr)\perp}\}$ as the only

relevant coupling parameters in the present effect. More explicitly, we identify the SOI in the material as the underlying mechanism in the effect, based on the corresponding expressions given below in **IV-3** and **Appendix C** for the parameters, which unambiguously indicates the SOI origin of $\{R_{SOI,eff}^{(4)\perp}, R_{SOI,eff}^{(4,corr)\perp}\}$.

**In-plane Zeeman effect due to $\vec{B}_{//}$**

For $\varepsilon = 0$ and $\mathbf{B} = B_x \hat{x}$, a coupling exists between opposite pseudospin states, which is given in the leading order by

$$\langle K, k_x \sim 0, n | H_{eff}^{(off-diag)} | K', k_x \sim 0, n \rangle = -E_{Z,eff}^{//}/2,$$

$$E_{Z,eff}^{//} \approx \mu_B B_x \left(\frac{\hbar^2 k_{y,n}^2}{2m^*}\right)$$

$$\left\{ -\frac{8}{\overline{\Delta}} \left(g_{eff}^{(3)//} - g_{eff}^{(2)//}\right) \frac{a \sin(KW_y)}{W_y} - \frac{16}{\overline{\Delta}} \left(g_{eff}^{(3)//} g_{eff}^{(3,corr)//}\right) \left(\frac{\sqrt{2m^* V_0}}{\hbar} a\right) \frac{a \cos(KW_y)}{W_y} \right\}$$

(26)

in the hard-wall limit. Due to the coupling, energy eigenstates in the subspace expanded by $\{|\tau, k_x, n\rangle$'s, $\tau = K, K'\}$ are given by $|+, k_x, n\rangle_x$ and $|-, k_x, n\rangle_x$ with a Zeeman-split subband structure, in agreement with the result shown in **Figure 3(b)**. Our theory yields a constant energy splitting "$E_{Z,eff}^{//}$" for states near $k_x = 0$. In addition, Eqn. (26) predicts an oscillatory and decaying behavior in the Zeeman energy splitting when increasing $W_y$, and the prediction is numerically confirmed by the same tight-binding calculation used to obtain **Figure 3(b)**.

Note that $\{g_{eff}^{(2)//}, g_{eff}^{(3)//}, g_{eff}^{(3,corr)//}\}$ are the only relevant coupling parameters in the present effect.

As the coupling parameters $\{g_{eff}^{\perp}, g_{eff}^{(2)//}, g_{eff}^{(3)//}, g_{eff}^{(3,corr)//}, R_{VOI,eff}, R_{SOI,eff}^{(4)\perp}, R_{SOI,eff}^{(4,corr)\perp}\}$ determine magnitudes of the various effects just discussed, order-of-magnitude expressions for them are relevant and presented below for reference:

$$g_{eff}^{\perp} = O\left[|P_{vc}|^2 / m_e \overline{\Delta}^2\right],$$

$$g_{eff}^{(2)//} = O\left[\frac{\overline{\Delta}}{\Delta_{so}} \overline{\lambda}^{(IR-flip)}\right] g_e,$$

$$g_{eff}^{(3)//} = O\left[\frac{\overline{\Delta}}{\Delta_{so}}\right] g_e,$$

$$g_{eff}^{(3,corr)//} = O\left[\overline{\lambda}^{(IR-flip)}\right],$$

$$R_{VOI,eff} = O\left[(e\hbar/m_e^2)|P_{vc}|^2/\overline{\Delta}^2\right],$$

$$R_{SOI,eff}^{(4)\perp} = O(\frac{\Delta_{so} e a^2}{\hbar \overline{\Delta}}),$$

$$R_{SOI,eff}^{(4,corr)\perp} = O\left[\frac{\overline{\Delta}}{\Delta_{so}} \overline{\lambda}^{(IR-flip)}\right].$$

(27)

($e$ = electron charge magnitude, $\Delta_{so}$ = spin-orbit gap parameter in valence band, $\overline{\Delta}$ = typical gap, $P_{vc}$ = momentum matrix element between conduction band minimum (*CBM*) and *VBM* states in the same valley, and $\overline{\lambda}^{(IR-flip)}$ is dimensionless and represents the typical coupling strength for simultaneous valley and IR index flipping relative to that for only valley flipping (see **Appendices B** and **D**).) Note that in the case of g-factor, Eqn. (27) yields $g_{eff}^{\perp} = O(1)$. For comparison, the experimental value is given by $g_{eff}^{\perp} \simeq 9$ [66].

With $H_{eff}$ completely specified above, a summary of symmetry properties of $H_{eff}$ is due here. In the case of quantum structures, it can be verified that $H_{eff}$ respects $T$ and $M_z$, if we ignore external fields. Moreover, if $U_{elastic}$ is taken to be an even function of $y$, it also respects $M_y$, in consistency with our choice of $x$-axis in the armchair direction.

**IV-2. Bare models**

We introduce below only "minimal" bare models essential for deriving primary parameter-dependent Hamiltonian terms in $H_{eff}$. **Appendix B** presents certain mathematical details of the models and also an extension that can generate secondary terms.

As illustrated earlier in **Figure 5**, quantum paths are divided into two classes - **Class A** of "IR-conserved" nature and **Class B** of "IR-flipped" nature. They will be identified and presented below for each configuration, according to the two following rules. Firstly, they contribute terms to $H_{eff}^{(off-diag)}$ up to the first order in momentum $\vec{\Pi}$. This rule is adopted based on the evidence given in **IV-1** that $H_{eff}^{(off-diag)}$ with such terms produce vertical Rashba and in-plane Zeeman effects in agreement with those in **III** derived with symmetry-based analysis. Secondly, they generate primary parameter-dependent terms in $H_{eff}^{(off-diag)}$, i.e., those involving $\{g_{eff}^{(2)//}, g_{eff}^{(3)//}, R_{SOI,eff}^{(4)\perp}\}$. The two rules define the scope of minimal models.

**Twelve-state model for the vertical configuration**

In the vertical configuration, the coupling between |K⟩ and |K'⟩ comes primarily from the four-step quantum paths consisting of 1) valley-flip scattering, 2) SOI-induced spin flipping, 3) $\vec{k} \cdot \vec{p}$ coupling, and 4) $\varepsilon_z$-induced parity mixing.

With an analysis based on permutation of the four steps, such paths involve ten intermediate states, with five characterized by ($D_0$, ↑, $K$), ($D_0$, ↓, $K$), ($D_{-2}$, ↓, $K$), ($D_{-1}$, ↑, $K$), or ($D_1$, ↓, $K$), in valley-$K$, and the rest by ($D_0$, ↑, $K'$), ($D_0$, ↓, $K'$), ($D_2$, ↑, $K'$), ($D_{-1}$, ↑, $K'$), or ($D_1$, ↓, $K'$), in valley-$K'$, and the paths can be totally captured by a twelve-state $\vec{k}\cdot\vec{p}$ model constructed in the following space, with basis states including the two $VBM$ states in addition to the intermediate ones, namely, $\{|\varphi_1\rangle = |VBM, K\rangle$, $|\varphi_2\rangle = |\Psi_{D_0}^{(n_1)}, \uparrow, K\rangle$, $|\varphi_3\rangle = |\Psi_{D_0}^{(n_1')}, \downarrow, K\rangle$, $|\varphi_4\rangle = |\Psi_{D_{-1}}^{(n_2)}, \uparrow, K\rangle$, $|\varphi_5\rangle = |\Psi_{D_1}^{(n_3)}, \downarrow, K\rangle$, $|\varphi_6\rangle = |\Psi_{D_{-2}}^{(n_4)}, \downarrow, K\rangle$, $|\varphi_7\rangle = |VBM, K'\rangle$, $|\varphi_8\rangle = |\Psi_{D_0}^{(n_1)}, \downarrow, K'\rangle$, $|\varphi_9\rangle = |\Psi_{D_0}^{(n_1')}, \uparrow, K'\rangle$, $|\varphi_{10}\rangle = |\Psi_{D_1}^{(n_2)}, \downarrow, K'\rangle$, $|\varphi_{11}\rangle = |\Psi_{D_{-1}}^{(n_3)}, \uparrow, K'\rangle$, $|\varphi_{12}\rangle = |\Psi_{D_2}^{(n_4)}, \uparrow, K'\rangle\}$. Superscripts "$n_1$" and etc. are representative band indices of intermediate states. Here, we have used the notations $|VBM, K\rangle$ and $|VBM, K'\rangle$ in place of $|K\rangle$ and $|K'\rangle$, respectively, to explicitly indicate their locations at the valence band maximum. $\varphi_7$-$\varphi_{12}$ are time reversal conjugates of $\varphi_1$-$\varphi_6$. This is essential to ensure that the model so constructed satisfies the $T$-symmetry, in the absence of any magnetic field. In the model, quantum paths for the $|K\rangle$-$|K'\rangle$ coupling are classified into four types, according to the intermediate states involved, as depicted in **Figure 6**. Corresponding contributions from them to the coupling are all given by fourth-order perturbation-theoretical expressions. In contrast, other contributions that involve intermediate states outside the twelve-state space, such as those of ($D_2$, $s_z = \pm1$, $K$) or ($D_{-2}$, $s_z = \pm1$, $K'$), are generally of higher order. An example is given below

$$(VBM, K) \xrightarrow{\varepsilon_z z} \left(\Psi_{D_{-1}}^{(n_2)}, \uparrow, K\right) \xrightarrow{\vec{L}\cdot\vec{s}} \left(\Psi_{D_0}^{(n_1)}, \downarrow, K\right) \xrightarrow{\vec{k}\cdot\vec{p}} \left(\Psi_{D_2}^{(v)}, \downarrow, K\right) \quad (28)$$
$$\xrightarrow{U_{elastic}} \left(\Psi_{D_2}^{(n_4)}, \downarrow, K'\right) \xrightarrow{\vec{k}\cdot\vec{p}} (VBM, K'),$$

which is fifth-order and $O((\hbar/m_e)\vec{k}\cdot\vec{p}/\overline{\Delta})$ smaller than leading, fourth-order ones ($\overline{\Delta} = O(eV)$).

The wave equation in the bare model is formulated in the effective-mass approximation. Let the wave function $|\Phi\rangle = \sum_i F_i(x,y)|\varphi_i\rangle$, where $F_i$'s are envelope functions. $F_i$'s satisfy the following Hamiltonian equation

$$\sum_j \left(H^\perp\right)_{ij} F_j = E F_i,$$
$$H^\perp = H_{band} + H_{\vec{k}\cdot\vec{p}} + H_{SOI,off-diag} + H_\varepsilon^\perp \quad (29)$$
$$+ H_Z^\perp + U_{diag} + U_{K\leftrightarrow K'}.$$

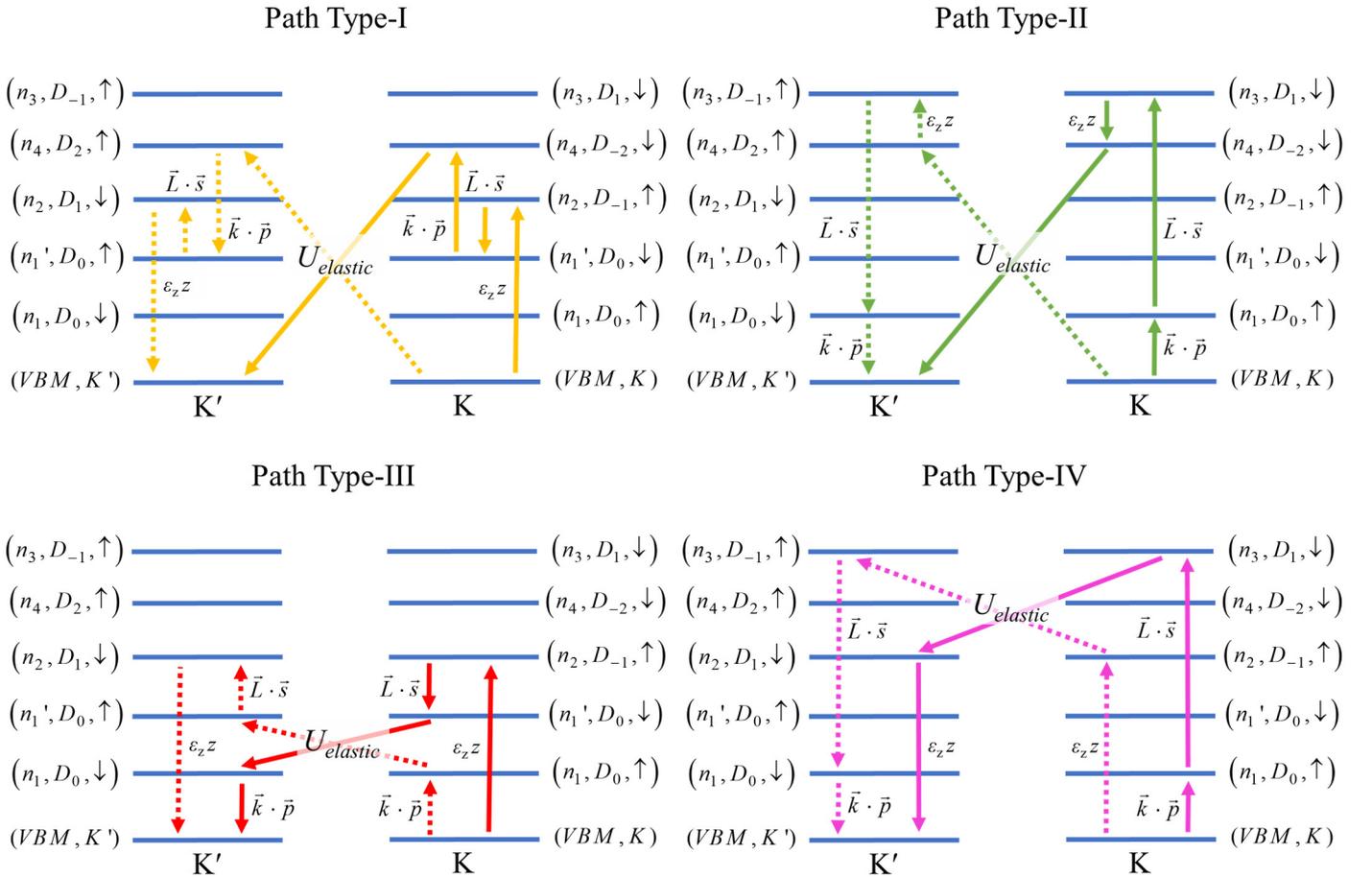

**Figure 6** Four types of quantum paths, all belonging to **Class A**, for $|K\rangle \to |K'\rangle$ in the vertical field configuration. They generate the $R_{SOI,eff}^{(4)\perp}$-dependent Hamiltonian term in $H_{eff}^{(off-diag)}$. Each type consists of two paths of same color but different line styles, one solid and the other dashed, with opposite operator sequences. $(n_1, D_0, \downarrow)$ and etc. denote intermediate states.

$H^\perp$ is the Hamiltonian. Matrix elements of the various terms in $H^\perp$ are given in **Appendix B**. $H_{band}$ describes the "bare" energy bands of an electron and consists of only diagonal matrix elements given by

$$(H_{band})_{ij} = \left(E_i + p^2/2m_i^\perp\right)\delta_{ij} \quad (30)$$

where $m_i^\perp$ is the "bare" mass and $E_i$ is the band edge energy for basis state $|\varphi_i\rangle$. $H_{\vec{k}\cdot\vec{p}}$ describes the $\vec{k}\cdot\vec{p}$ coupling "$\frac{\hbar}{m_e}\vec{k}\cdot\vec{p}$". We divide the SOI into $H_{SOI,off\text{-}diag}$ - the SOI-induced spin flipping and the diagonal part, with the latter shifting the band edge and merged into $H_{band}$. $H_\varepsilon^\perp$ describes $\varepsilon_z$-induced parity-mixing between states. $H_Z^\perp$ describes the Zeeman interaction due to $B_z$.

$U_{diag}$ and $U_{K\leftrightarrow K'}$ both derive from $U_{elastic}$. See **Appendix D** for a detailed discussion. $U_{diag}$ is the valley-conserving part, with

$$(U_{diag})_{ij} = \eta_i U_{elastic}\delta_{ij}. \quad (31)$$

where $\eta_i$ is the relative potential strength for basis state $|\varphi_i\rangle$, For a bulk, $\eta_i = 1$ for all states. For quantum structures, $\eta_1 = \eta_7 = 1$ for the valence band states, and $\eta_{i\neq 1,7}$ depends on band offsets. It differs from unity if band offsets result in a difference in the quantum confinement potentials for $|\varphi_1\rangle$ and $|\varphi_i\rangle$. $U_{K\leftrightarrow K'}$ is the valley-mixing part given by

$$(U_{K\leftrightarrow K'})_{i\leq 6, j\geq 7} = (U_{K\leftrightarrow K'})_{j\geq 7, i\leq 6}^* $$
$$= \lambda_{i,j} U_{elastic}^{(derived)}(\vec{r}; G_i, G_j) e^{-2i\vec{K}\cdot\vec{r}}. \quad (32)$$

$U_{elastic}^{(derived)}$ is a potential energy function derived from $U_{elastic}$. $G_i$ and $G_j$ are the irreducible representation indices of $|\varphi_i\rangle$ and $|\varphi_j\rangle$, respectively. $\lambda_{i,j}$ is a relative, dimensionless strength parameter which depends on state indices. In particular, $\lambda_{i,j}$ is both spin and parity diagonal, since $U_{elastic}^{(derived)}$ is nonmagnetic and even in $z$, the same as $U_{elastic}$. We provide $U_{elastic}^{(derived)}$ below in a few cases of interest.

(i) In the case of a bulk with dilute, random distribution of identical, short-range impurities on the M-sublattice,

$$U_{elastic}^{(derived)}(\vec{r}; G_i, G_j) = \sum_{R_i} v_{impurity}(0)\delta_{\vec{R}(\vec{r}),\vec{R}_i}. \quad (33)$$

(ii) In the case of quantum structures,

$$U_{elastic}^{(derived)}(\vec{r}; G_i = G_j) \approx U_{elastic}(\vec{r}); \quad (34)$$

$$U_{elastic}^{(derived)}(\vec{r}; G_i \neq G_j) \approx a(\partial_x, \partial_y)_{-\text{sgn}(G_i, G_j)} U_{elastic}(\vec{r}). \quad (35)$$

Above, $\text{sgn}(G_i, G_j) = -\text{sgn}(G_j, G_i) = +$, for $(G_i, G_j) = (D_0, D_2)$, $(D_2, D_{-2})$, and $(D_{-1}, D_1)$, and $\text{sgn}(G_i, G_j) = 0$ otherwise. $(\partial_x, \partial_y)_0 = 0$ and $(\partial_x, \partial_y)_\pm = \partial_x \pm i\partial_y$.

**Eight-state model for the in-plane configuration**

In the presence of an in-plane magnetic field $\vec{B}_{//}$, the coupling between $|K\rangle$ and $|K'\rangle$ comes from two-step quantum paths consisting of 1) elastic scattering and 2) magnetic field-induced spin flipping, or three-step ones consisting of 1) elastic scattering, 2) magnetic field-induced spin flipping, and 3) $\vec{k}\cdot\vec{p}$ coupling. Corresponding quantum paths are classified into four types as depicted in **Figure 7**, with contributions to the theoretical expressions involving intermediate states of ($D_{\pm 2}$, $s_z = \pm 1$, $\tau = K, K'$). Quantum paths using other intermediate states make higher-order contributions, such as the example given below

$$(VBM, K) \xrightarrow{\vec{k}\cdot\vec{p}} \left(\Psi_{D_0}^{(n')}, \uparrow, K\right) \xrightarrow{\vec{s}_{//}\cdot\vec{B}_{//}} \left(\Psi_{D_0}^{(n')}, \downarrow, K\right)$$
$$\xrightarrow{U_{elastic}} \left(\Psi_{D_0}^{(n)}, \downarrow, K'\right) \xrightarrow{\vec{k}\cdot\vec{p}} (VBM, K'), \quad (36)$$

which is fourth order and $O((\hbar/m_e)\vec{k}\cdot\vec{p}/\bar{\Delta})$ smaller than third-order ones ($\bar{\Delta} = O(eV)$).

For the in-plane configuration, an eight-state $\vec{k}\cdot\vec{p}$ model is constructed in the space with basis states $\{|\varphi_1\rangle = |VBM, K\rangle$, $|\varphi_2\rangle = \left|\Psi_{D_2}^{(v)}, \downarrow, K\right\rangle$, $|\varphi_3\rangle = \left|\Psi_{D_{-2}}^{(n)}, \uparrow, K\right\rangle$, $|\varphi_4\rangle = \left|\Psi_{D_{-2}}^{(n)}, \downarrow, K\right\rangle$, $|\varphi_5\rangle = |VBM, K'\rangle$, $|\varphi_6\rangle = \left|\Psi_{D_{-2}}^{(v)}, \uparrow, K'\right\rangle$, $|\varphi_7\rangle = \left|\Psi_{D_2}^{(n)}, \downarrow, K'\right\rangle$, $|\varphi_8\rangle = \left|\Psi_{D_2}^{(n)}, \uparrow, K'\right\rangle\}$, where $\varphi_5$-$\varphi_8$ are time reversal conjugates of $\varphi_1$-$\varphi_4$.

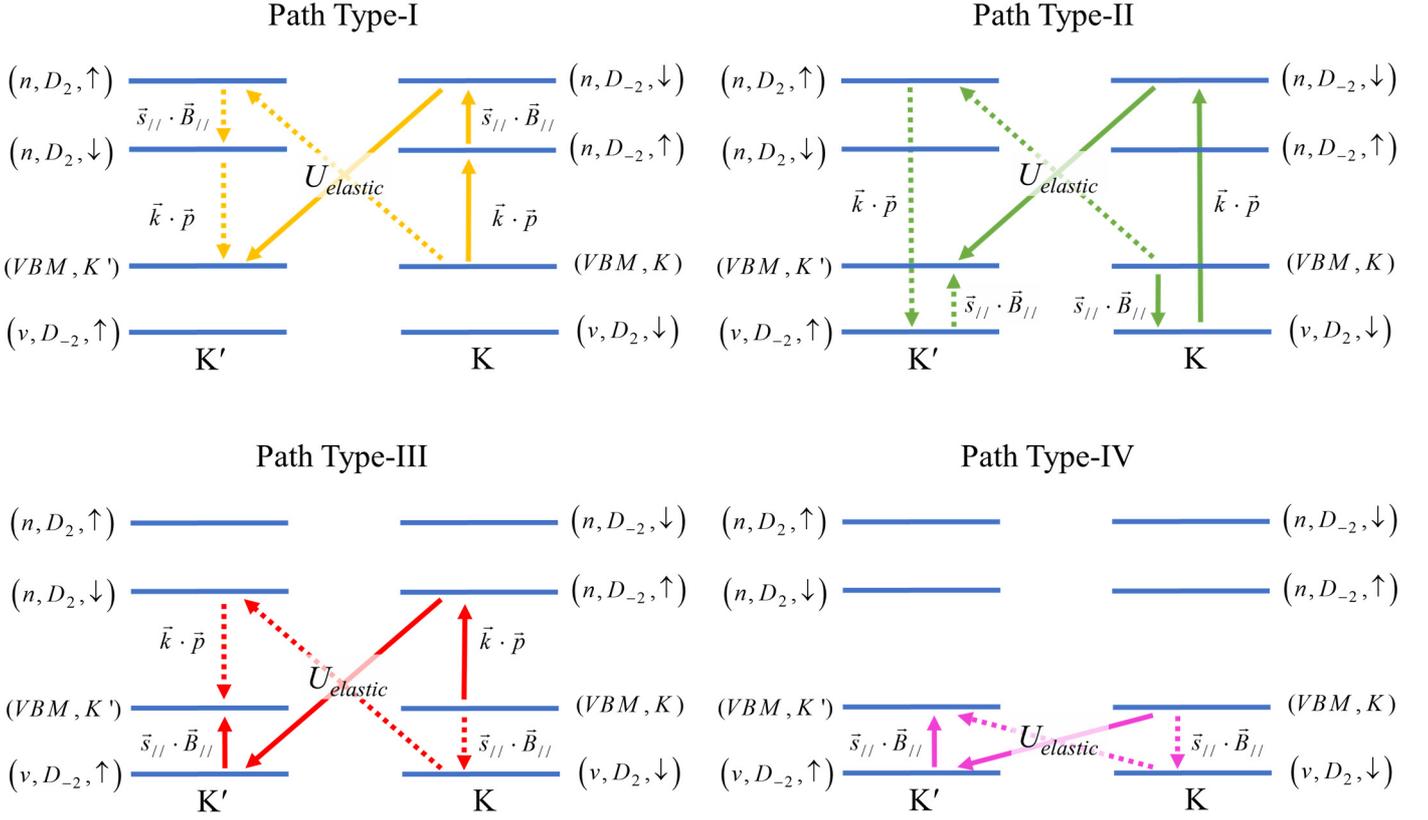

**Figure 7** Four types of quantum paths for $|K\rangle \to |K'\rangle$ in the in-plane-field configuration, with Type- I, II, and III in **Class A** and IV in **Class B**. Class A paths generate the $g_{eff}^{(3)//}$-dependent Hamiltonian term while **Class B** paths generate the $g_{eff}^{(2)//}$-dependent term, in $H_{eff}^{(off-diag)}$.

The corresponding bare Hamiltonian is described below:

$$H^{//} = H_{band} + H_{\vec{k}\cdot\vec{p}} + H_{SOI,off-diag} + U_{diag} \quad (37)$$
$$+ U_{K\leftrightarrow K'} + H_{VOI}^{//} + H_{\varepsilon}^{//} + H_{Z}^{//}.$$

The first five terms on the right hand side have the same interpretations as their corresponding parts in $H^{\perp}$. $H_{VOI}^{//}$ is the VOI due to $\vec{\varepsilon}_{//}$, and $H_{\varepsilon}^{//}$ is the electric potential energy due to $\vec{\varepsilon}_{//}$. $H_Z^{//}$ describes the Zeeman interaction due to $\vec{B}_{//}$. We ignore the Landau orbital quantization in view of its significant suppression by the vertical confinement in a 2D layer. Detailed matrix elements of the various terms in $H^{//}$ are given in **Appendix B**.

### IV-3. Effective coupling parameters

A Schrieffer-Wolff transformation is performed on both models, reducing them to corresponding effective theories in the small space expanded by $\{|K\rangle, |K'\rangle\}$. See **Appendix C**. The reduction obtains coupling parameters in the effective theory in terms of both band structure and "bare coupling" parameters, providing an important revelation to the connection between the SVO physics and underlying band structure.

From the reduction of twelve-state model,

$$R_{SOI,eff}^{(4)\perp} = \sqrt{\frac{3}{8}}\bar{\Delta}\left(\frac{-i}{m_e}\right)$$

$$\left\{ \left[ \sum_{\substack{n_1' \in D_0, \\ n_2 \in D_{-1}, \\ n_4 \in D_{-2}}} \frac{\lambda_{v,n_4}\Delta_{so}^{(n_1',n_2)}(e\zeta_{n_2 v})P_{n_4 n_1'}}{(E_{VBM}-E_{n_2,\uparrow,K})(E_{VBM}-E_{n_1',\downarrow,K})(E_{VBM}-E_{n_4,\downarrow,K})} \right] \right.$$

$$+ \left[ \sum_{\substack{n_1 \in D_0, \\ n_3 \in D_1, \\ n_4 \in D_{-2}}} \frac{\lambda_{v,n_4}\Delta_{so}^{(n_3,n_1)}(e\zeta_{n_4 n_3})P_{n_1 v}}{(E_{VBM}-E_{n_1,\uparrow,K})(E_{VBM}-E_{n_3,\downarrow,K})(E_{VBM}-E_{n_4,\downarrow,K})} \right]$$

$$+ \left[ \sum_{\substack{n_1,n_1' \in D_0, \\ n_2 \in D_{-1}}} \frac{\lambda_{n_1',n_1}\Delta_{so}^{(n_1',n_2)}(-e\zeta_{n_2 v})P_{n_1 v}}{(E_{VBM}-E_{n_1,\uparrow,K})(E_{VBM}-E_{n_2,\uparrow,K})(E_{VBM}-E_{n_1',\downarrow,K})} \right]$$

$$+ \left[ \sum_{\substack{n_1 \in D_0, \\ n_2 \in D_{-1}, \\ n_3 \in D_1}} \frac{\lambda_{n_2,n_3} \Delta_{so}^{(n_3,n_1)} (-e\zeta_{n_2 v}) P_{n_1 v}}{(E_{VBM} - E_{n_2,\uparrow,K})(E_{VBM} - E_{n_3,\downarrow,K})(E_{VBM} - E_{n_1,\uparrow,K})} \right] \right\},$$

(38)

$$g_{eff}^{\perp} = g_e + g_{valley-orbital},$$

$$g_{valley-orbital} = \frac{1}{m_e} \left[ \sum_{l \in D_{-2}} \frac{|P_{vl}|^2}{(E_{VBM} - E_{l,\uparrow,K})} \right. \quad (39)$$

$$\left. - \sum_{l \in D_0} \frac{|P_{vl}|^2}{(E_{VBM} - E_{l,\uparrow,K})} \right].$$

$\Delta_{so}^{(n_1',n_2)}$ and etc. are SOI strength parameters. $\zeta_{n_2 v}$ and etc. are matrix elements of "$z$". For example, $\zeta_{n_2 v} = \langle \Psi_{D_{-1}}^{(n_2)}, \uparrow, K | z | VBM, K \rangle$. $P_{n_4 n_1}$ and etc. are momentum matrix elements. Above coupling parameter expressions have been summed over states with the same irreducible representations as $\varphi_2 - \varphi_{11}$, in order to account for all leading-order contributions.

From the reduction of eight-state model,

$R_{VOI,eff}$

$$= \frac{e}{2\hbar}\left(\frac{\hbar}{m_e}\right)^2 \left[ \sum_{l \in D_0} \frac{|P_{vl}|^2}{(E_{VBM} - E_{l,\uparrow,K})^2} - \sum_{n \in D_{-2}} \frac{|P_{vn}|^2}{(E_{VBM} - E_{n,\uparrow,K})^2} \right],$$

(40)

$$g_{eff}^{(2)//} = \frac{\overline{\Delta}_{vv} g_e}{(E_{VBM} - E_{v,\downarrow,K})}.$$

(41)

$g_{eff}^{(3)//}$

$$= -i \frac{g_e \hbar \overline{\Delta}}{2 m_e a} \sum_{n \in D_{-2}} \lambda_{v,n} P_{nv} \left[ \frac{1}{(E_{VBM} - E_{n,\uparrow,K})(E_{VBM} - E_{n,\downarrow,K})} \right.$$

$$+ \frac{1}{\left[(E_{VBM} - E_{n,\downarrow,K})(E_{VBM} - E_{v,\downarrow,K})\right]}$$

$$\left. + \frac{1}{(E_{VBM} - E_{v,\downarrow,K})(E_{VBM} - E_{n,\uparrow,K})} \right]$$

(42)

Expressions for the secondary parameters $\{R_{SOI,eff}^{(4,corr)\perp}, g_{eff}^{(3,corr)//}\}$ are provided in **Appendix C**.

## V. SPIN-VALLEY-ORBITAL QUANTUM COMPUTING

Spin-valley-orbital quantum computing is proposed here with QD-confined holes as qubits. In such scheme the qubit state space is expanded by the Kramers pair of QD ground states, one labeled as $|K>_{QD}$ or $|K, m = 0, n = 0>$ with $l_z = 2$, $s_z = 1$ and the other $|K'>_{QD}$ or $|K', m = 0, n = 0>$ with $l_z = -2$, $s_z = -1$. "$m$" and "$n$" refer to the quantum labels for hole confinement in x- and y- directions, respectively.

Qubit states and all-electrical manipulation are discussed in **V-1** for the vertical configuration where $\boldsymbol{\varepsilon} = \varepsilon_z \hat{z}$ and $\mathbf{B} = B_z \hat{z}$, and in **V-2** for the in-plane configuration where $\boldsymbol{\varepsilon} = \varepsilon_y \hat{y}$ and $\mathbf{B} = B_x \hat{x}$. In **V-3**, we compare manipulation rates in the two configurations. In **V-4**, we briefly remark on issues of qubit initialization, readout, and qugates in the scheme.

**V-1. The vertical configuration**

The physics of qubits in this configuration is controlled by three Hamiltonian terms, as summarized below: 1) the potential energy "$U_{elastic}$" (= $U_{QD}$) in $H_{eff}^{(diag)}$ confines the carrier and determines qubit states; 2) the $B_z$-induced vertical Zeeman term "$\left(E_{Z,eff}^{\perp}/2\right)v_z$" in $H_{eff}^{(diag)}$ generates a Larmor precession in the Bloch sphere representation around the "z-axis going through $|K>_{QD}$ and $|K'>_{QD}$", providing one type of qubit manipulation; and 3) the $\varepsilon_z$-induced vertical Rashba term $H_{eff}^{(off-diag)} = -\alpha^{\perp} k_x v_y$ provides another type of manipulation – a rotation around the "y-axis" of Bloch sphere. 2) and 3) combined together accomplish an arbitrary qubit manipulation.

**Qubit states**

Let $\delta_{QW}$ = QW quantization energy (in the y-direction), and $\delta_{QD}$ = QD quantization energy = $\min(\delta_{QW}, \hbar \omega_x)$. The analysis below is performed in the regime where $\delta_{QD} \gg E_{Z,eff}^{\perp} \gg \|H_{eff}^{(off-diag)}\|^2 / \delta_{QD}$ ($\|...\|$ = norm), in the framework of perturbation theory, with $H_{eff}^{(off-diag)}$ the perturbation. The eigenstates of $H_{eff}^{(diag)}$ in the hard-wall approximation for $U_{QW}(y)$ are approximately given by

$$|K, m, n\rangle = X_m(x) Y_n(y) | VBM, K >,$$
$$|K', m, n\rangle = X_m(x) Y_n(y) | VBM, K' >,$$
$$X_m(x) = \text{harmonic oscillator wavefunction}, \quad (43a)$$
$$Y_n(y) = (2/W_y)^{1/2} \begin{cases} \cos(k_{y,n} y), n = 2n' \\ \sin(k_{y,n} y), n = 2n'+1 \end{cases},$$
$$m, n = 0, 1, 2, .......$$

with corresponding energy levels

$$E_{\tau,m,n} = E_{m,n} + \tau E_{Z,eff}^{\perp}/2,$$

$$E_{m,n} = (\hbar k_{y,n})^2/2m^* + \hbar\omega_x (m + 1/2), \qquad (43b)$$

The ground states $|K,0,0\rangle$ and $|K',0,0\rangle$ expand the qubit state space. In writing above eigenstates, we have neglected the Landau orbital effect and made the replacement $\vec{\Pi} \to \vec{p}$ in $H_{eff}^{(diag)}$ due to two considerations. Firstly, we work within the regime where the QD confinement dominates over the Landau orbital confinement. Secondly, the magnetic field is primarily introduced to provide the Larmor precession for qubit manipulation. As will be shown below, the manipulation rate obtained in the present approximation scales with the Zeeman energy $E_{Z,eff}^{\perp}$ in the leading order. Inclusion of the Landau orbital effect here would only produce the next-order correction in the discussion of manipulation.

Next, we discuss the effect of $H_{eff}^{(off-diag)}$ for qubit manipulation. In the hard-wall approximation, we obtain

$$\langle K|H_{eff}^{(off-diag)}|K'\rangle_{QD} \simeq i\alpha^{\perp}|_{n=0}\langle k_x\rangle_{QD} + O(B_z), \qquad (44)$$

where $\langle k_x\rangle_{QD} = -i\int_{-\infty}^{\infty} X_0^*(x)\partial_x X_0(x)$.

Eqn. (44) shows that the mixing between $|K\rangle_{QD}$ and $|K'\rangle_{QD}$ scales, in the limit of weak $B_z$, with $\langle k_x\rangle_{QD}$. This result has two implications. Firstly, it vanishes since $\langle k_x\rangle_{QD} \propto \frac{d\langle x\rangle}{dt}=0$ for an energy eigenstate due to the Ehrenfest theorem, indicating a protection for the state from pseudospin flipping. Secondly, when a pseudospin flipping manipulation is intended, it suggests the application of an ac auxiliary electric field in the x-direction, which can generate a finite $\frac{d\langle x\rangle}{dt}$ as discussed next.

**Larmor precession, Rabi oscillation, and qubit manipulation**

In the qubit state space, the Hamiltonian in the leading order is given by (with $E_{0,0}$ omitted from the diagonal terms)

$$H_{qubit}^{\perp} \simeq \text{sgn}\left(E_{Z,eff}^{\perp}\right)\frac{\hbar\omega_L^{\perp}}{2}v_z - \alpha^{\perp}|_{n=0}\langle k_x\rangle_{QD} v_y, \qquad (45)$$

where $\omega_L^{\perp}$ is the Larmor frequency given by

$$\omega_L^{\perp} = |E_{Z,eff}^{\perp}|/\hbar. \qquad (46)$$

Next, consider the application of an ac in-plane electric field $\varepsilon_{ac}\cos(\omega_{ac}t)$ in the adiabatic regime where $\hbar\omega_{ac} \ll \delta_{QD}$. We provide a relatively intuitive discussion within the adiabatic approximation [67] for this regime. In the ac field, the total QD confinement potential in x-direction becomes time-dependent, with the center $x_0(t)$ being oscillatory:

$$U_{quad}(x) + e\varepsilon_{ac}x\cos(\omega_{ac}t)$$
$$= (1/2)m^*\omega_x^2[x - x_0(t)]^2 + O(\varepsilon_{ac})^2, \qquad (47)$$
$$x_0(t) = -e\varepsilon_{ac}\cos(\omega_{ac}t)/m^*\omega_x^2,$$

correct up to $O(\varepsilon_{ac})$. Within the adiabatic approximation, it results in the following dynamical qubit state, namely, a harmonic oscillator ground state with wave function centering around $x_0(t)$. This leads to

$$\langle k_x\rangle_{QD} \simeq \frac{m^*}{\hbar}dx_0/dt. \qquad (48)$$

**Appendix C** provides an alternative derivation with the Schrieffer-Wolff reduction.

The type of Hamiltonian in Eqn. (45) along with Eqn. (48) constitutes the well-known problem, namely, a two-state system with ac field-driven inter-state coupling.[68] Consider the case where $\text{sgn}(E_{Z,eff}^{\perp}) > 0$ and $\text{sgn}(\alpha^{\perp}|_{n=0}) > 0$. For $\omega_{ac} = \omega_L^{\perp}$, the standard rotating wave approximation (RWA) yields

$$H_{qubit}^{\perp} \sim \hbar\begin{pmatrix} \omega_L^{\perp}/2 & -\Omega_R^{\perp}e^{-i\omega_{ac}t} \\ -\Omega_R^{\perp}e^{i\omega_{ac}t} & -\omega_L^{\perp}/2 \end{pmatrix}\Bigg|_{\omega_{ac}=\omega_L^{\perp}}, \qquad (49)$$

$$\Omega_R^{\perp} = |\alpha^{\perp}|_{n=0}|e\varepsilon_{ac}\omega_{ac}/2\hbar^2\omega_x^2, \qquad (50)$$

and the corresponding time-dependent wave solution describes a Rabi oscillation between states $|K\rangle_{QD}$ and $|K'\rangle_{QD}$ ($\Omega_R^{\perp}$ = Rabi frequency). In the case where the initial state $\psi^{\perp}(0) = |K\rangle_{QD}$ for example, it gives

$$\psi^{\perp}(t) = c_K(t)|K\rangle_{QD} + c_{K'}(t)|K'\rangle_{QD},$$
$$c_K(t) = \cos(\Omega_R^{\perp}t), \quad c_{K'}(t) = i\sin(\Omega_R^{\perp}t), \qquad (51)$$

in the rotating reference frame.

Let $W_x = 1.5$ $W_y = 15a$, $V_0 = 1$ $eV$, electric fields $\varepsilon_{ac} = 0.4$ $mV/a$, $\varepsilon_z = 10$ $mV/a$, and magnetic field $B_z \sim 0.2$ $T$. Using $g_{eff}^{\perp} = 9$, we have $E_{Z,eff}^{\perp} = 0.11$ $meV$. In the case of $WSe_2$, with $m^* = 0.36$ $m_e$ [46], it gives $\delta_{QW} \sim 98$ $meV$, $\hbar\omega_x \sim 8.9$ $meV$, $\alpha^{\perp} \sim 0.18$ $meV\cdot a$, and $\Omega_R^{\perp} \sim 72$ $MHz$. In the case of $MoSe_2$, with $m^* = 0.6$ $m_e$ [46], we have $\delta_{QW} \sim 59$ $meV$, $\hbar\omega_x \sim 5.4$ $meV$, $\alpha^{\perp} \sim 0.032$ $meV\cdot a$, and $\Omega_R^{\perp} \sim 35$ $MHz$, due to a weaker SOI.

**V-2. The in-plane configuration**

A close analogy exists between the qubit physics here and that in the vertical configuration. In particular, 1) $B_x$ induces an in-plane Zeeman effect, by which the pseudospin is quantized into states denoted below as $|+\rangle_{QD}$ and $|-\rangle_{QD}$ (symmetric and antisymmetric combinations of $|K>_{QD}$ and $|K'>_{QD}$, respectively), with the Zeeman energy splitting $E_{Z,eff}^{//}$ between them. Such splitting generates a precession around the "x-axis going through $|+\rangle_{QD}$ and $|-\rangle_{QD}$" in the Bloch sphere; and 2) $\varepsilon_y$ induces an in-plane Rashba effect producing a coupling "$\alpha^{//}\langle k_x\rangle_{QD}$" between $|+\rangle_{QD}$ and $|-\rangle_{QD}$, which enables, in the presence of an ac electric field in the x-direction, a rotation around the "z-axis".

**Qubit states**

We perform a perturbation-theoretical analysis in the regime where $\delta_{QW} \gg \|e\varepsilon_y y\|$ and $\delta_{QD} \gg E_{Z,eff}^{//} \gg \|H_{VOI}^{(val)}\|^2/\delta_{QD}$. Consider $H_{eff}^{(diag)}\big|_{\varepsilon_y=0}$ first. Eigenstates of $H_{eff}^{(diag)}\big|_{\varepsilon_y=0}$ are given by $\{|\tau,m,n\rangle\text{'s},\ \tau = K, K'\}$, the same as those in the vertical case, but with energy levels given by

$$E_{m,n} = (\hbar k_{y,n})^2/2m^* + \hbar\omega_x(m+1/2) \quad (52)$$

without the Zeeman term. The ground states $|K>_{QD}$ and $|K'>_{QD}$ again expand the qubit state space. Next, consider effects of the terms ignored, in the qubit state space. Specifically, with

$$\langle K|e\varepsilon_y y|K\rangle_{QD} = \langle K'|e\varepsilon_y y|K'\rangle_{QD} = 0,$$
$$\langle K|H_{VOI}^{(val)}|K\rangle_{QD} = -\langle K'|H_{VOI}^{(val)}|K'\rangle_{QD} = \alpha^{//}\langle k_x\rangle_{QD},$$
$$\langle K|H_{eff}^{(off-diag)}|K'\rangle_{QD} \simeq -E_{Z,eff}^{//}/2, \quad (53)$$

it gives the following qubit Hamiltonian (with $E_{0,0}$ omitted from the energy terms)

$$H_{qubit}^{//} \simeq -\text{sgn}(E_{Z,eff}^{//})\frac{\hbar\omega_L^{//}}{2}v_x + \alpha^{//}\langle k_x\rangle_{QD}v_z, \quad (54)$$

up to the first order of $\varepsilon_y$ and $B_x$. Here, the Larmor frequency $\omega_L^{//} = |E_{Z,eff}^{//}|/\hbar$. Since $\langle k_x\rangle_{QD} = 0$, eigenstates of $H_{qubit}^{//}$ are given by $|+\rangle_{QD}$ and $|-\rangle_{QD}$, with

$$|\pm\rangle_{QD} = (|K\rangle_{QD} \pm |K'\rangle_{QD})/\sqrt{2}, \quad (55)$$

which are split by the Zeeman energy $E_{Z,eff}^{//}$.

**Larmor precession, Rabi oscillation, and qubit manipulation**

For qubit manipulation, an ac-electric field in the x-direction, $\varepsilon_{ac}\cos(\omega_{ac}t)$, is introduced. In the adiabatic approximation, we make the substitution $<k_x>_{QD} \to \frac{m^*}{\hbar}dx_0/dt$, and Eqn. (44) becomes

$$H_{qubit}^{//} \simeq -\text{sgn}(E_{Z,eff}^{//})\frac{\hbar\omega_L^{//}}{2}v_z + 2\text{sgn}(\alpha^{//})\hbar\Omega_R^{//}\sin(\omega_{ac}t)v_x, \quad (56)$$

in the basis of $\{|+\rangle_{QD}, |-\rangle_{QD}\}$, where $\Omega_R^{//}$ is the Rabi frequency given by

$$\Omega_R^{//} = |\alpha^{//}|e\varepsilon_{ac}\omega_{ac}/\hbar^2\omega_x^2. \quad (57)$$

Eqn. (56) can also be derived with the Schrieffer-Wolff reduction in **Appendix C**.

Consider the case where $\text{sgn}(E_Z^{//}) > 0$ and $\text{sgn}(\alpha^{//}) > 0$. For $\omega_{ac} = \omega_L^{//}$, the RWA yields

$$H_{qubit}^{//} \sim \hbar\begin{pmatrix} -\omega_L^{//}/2 & -i\Omega_R^{//}e^{i\omega_{ac}t} \\ i\Omega_R^{//}e^{-i\omega_{ac}t} & \omega_L^{//}/2 \end{pmatrix}, \quad (58)$$

and the corresponding wave solution is, in the case where the initial state $\psi^{//}(t=0) = |+\rangle_{QD}$ for example, given by

$$\psi^{//}(t) = c_+(t)|+\rangle_{QD} + c_-(t)|-\rangle_{QD},$$
$$c_+(t) = \cos(\Omega_R^{//}t),\ c_-(t) = \sin(\Omega_R^{//}t), \quad (59)$$

in the rotating reference frame.

Let $W_x = 1.5\ W_y = 15a$, $V_0 = 1\ eV$, electric fields $\varepsilon_{ac} = 0.4\ mV/a$, $\varepsilon_y = 5\ mV/a$ and magnetic field $B_x = 1\ T$. It gives $\alpha^{//} \sim 2.5\ meV\cdot a$. In the case of MoSe$_2$, we obtain $E_Z^{//} \sim 30\ \mu eV$ and $\Omega_R^{//} \sim 1.5\ GHz$. For WSe$_2$, due to a stronger SOI, it gives $E_Z^{//} \sim 20\ \mu eV$ and $\Omega_R^{//} \sim 370\ MHz$.

Results in **V-1** and **V-2** are summarized in **Figure 8**, which shows the time evolution of qubit states in the Bloch sphere, in both the lab and rotating reference frames.

**V-3. Comparison between configurations**

We compare manipulation rates in the two configurations. In the resonance condition where $\omega_{ac} = \omega_L^{\perp}$ in Eqn. (50) and $\omega_{ac} = \omega_L^{//}$ in Eqn. (57), it shows that $\Omega_R^{\perp}(\Omega_R^{//})$ is dependent on the $|K>_{QD}$-$|K'>_{QD}$ ($|+>_{QD}$-$|->_{QD}$) coupling strength, the ac electric field strength, and the $|K>_{QD}$-$|K'>_{QD}$ ($|+>_{QD}$-$|->_{QD}$) energy splitting. Therefore, under the same ac electric field, we obtain the following ratio

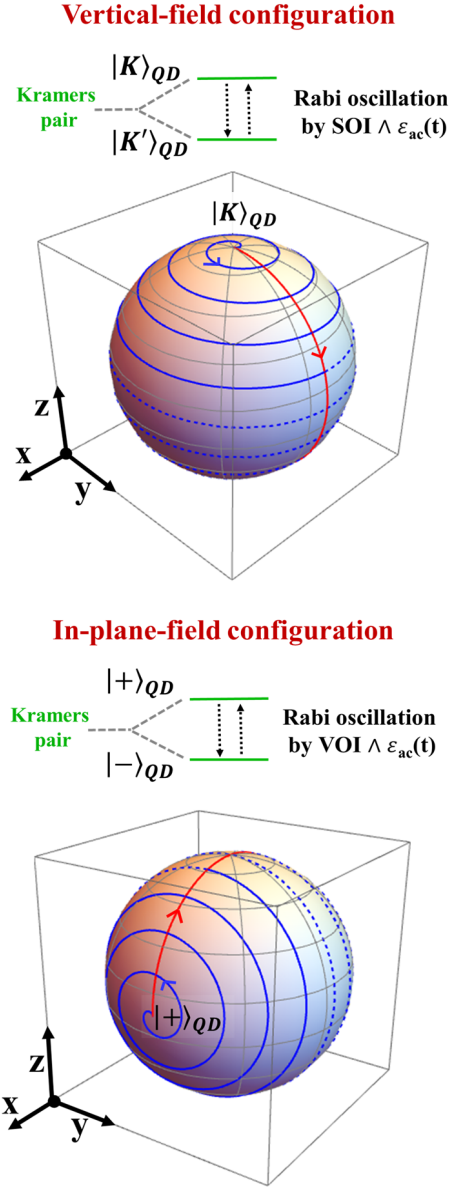

**Figure 8** Qubit state evolution on the Bloch sphere, in both vertical and in-plane field configurations. It shows $|K\rangle_{QD} \to |K'\rangle_{QD}$ ($|+\rangle_{QD} \to |-\rangle_{QD}$) in the vertical (in-plane) configuration, viewed in the lab reference frame (blue cure) and in the rotating reference frame (red curve). For the plot, we use $\omega_L^\perp = 40\Omega_R^\perp$ and $\omega_L^{//} = 40\Omega_R^{//}$. Upper graphs depict corresponding transitions between qubit states, which are effected by ac electric field-induced Rabi oscillations, based on the SOI and VOI mechanisms in the vertical and in-plane configurations, respectively.

$$\Omega_R^\perp / \Omega_R^{//}$$
$$\sim \left|\frac{\alpha^\perp}{\alpha^{//}}\right| \frac{E_{Z,eff}^\perp}{E_{Z,eff}^{//}} \qquad (60)$$
$$\sim \frac{\hbar}{a\sqrt{m^* V_0}} \left(\frac{g_{eff}^\perp}{g_{eff}^{//}}\right) \left(\frac{B_z}{B_x}\right) \left(\frac{\varepsilon_z R_{SOI,eff}^{(4)\perp}}{\varepsilon_y R_{VOI,eff}}\right)$$

Above, $g_{eff}^{//}$ denotes O($g_{eff}^{(2)//}$) (= O($g_{eff}^{(3)//}$)), and the contributions in $\Omega_R^\perp$ and $\Omega_R^{//}$ involving secondary coupling parameters have been ignored. Two points are noted below based on Eqn. (60). Firstly, since the vertical configuration depends on the SOI for the manipulation, $\Omega_R^\perp \propto R_{SOI,eff}^{(4)\perp} \propto \Delta_{so}$, which favors W-based TMDCs over Mo-based ones. Secondly, the in-plane configuration attempts to quantize the pseudospin "in the plane" for the manipulation. Therefore, it has to overcome the SOI that quantizes the spin in the out-of-plane direction. This results in $\Omega_R^{//} \propto g_{eff}^{//} \propto \Delta_{so}^{-1}$, in favor of Mo-based TMDCs over W-based ones.

We also note a few points in the numerical estimation of Rabi frequencies given earlier. Firstly, the dc electric field strengths there were chosen to be as large as possible in order to obtain favorable Rabi frequencies while at the same time it does not invalidate in a qualitative way the theoretical analysis presented. For example, while trying to optimize $\varepsilon_z R_{SOI,eff}^{(4)\perp}$ in the vertical configuration, a conservative $\varepsilon_z = 10\ mV/a$ was used which makes "$e\varepsilon_z a$" two orders of magnitude below atomic energy level spacing (~O($eV$)), in order to avoid a strong $\varepsilon_z$-induced atomic orbital mixing. On the other hand, $\varepsilon_y = 5\ mV/a$ was taken in order to maximize $\varepsilon_y R_{VOI,eff}$ in the in-plane configuration. In fact, at $\varepsilon_y = 5\ mV/a$, the corresponding potential energy across the QD, "$e\varepsilon_y W_y$", is comparable to the quantization energy in the $y$-direction, and the quantum state wave function $Y_n(y)$ may be modified quantitatively if not qualitatively. However, we do not expect such modification to affect the order of magnitude of Rabi frequencies estimated. In passing, we note that both the vertical and in-plane electric field strengths envisioned here are experimentally accessible. In particular, stronger vertical and in-plane field strengths at 200 $mV/a_{BLG}$ ($a_{BLG}$ = interlayer spacing in AB-stacked bilayer graphene) [69] and 10 $mV/Å$ [70], respectively, have been experimentally demonstrated. Secondly, at the above field strengths, we have $\varepsilon_z R_{SOI,eff}^{(4)\perp} / \varepsilon_y R_{VOI} = O(1)$, implying comparable SOI and VOI effects. Thirdly, $B_z \sim 0.2\ T$ and $B_x \sim 1\ T$ were used in the estimation based on the following experimental consideration. For $B_z \sim 0.2\ T$, the corresponding Larmor frequency $\omega_L^\perp = O(10^2\ GHz)$ already approaches the somewhat challenging radio frequency range for electrical signal processing. On the other hand, facilities to generate a magnetic field $\sim 1\ T$ are available in a number of labs. Overall, our estimation yields an optimized manipulation time $\sim \left(\Omega_R^{//}\right)^{-1} \sim$ O($ns$), which is comparable to that in the spin qubit case [71] and much shorter than the decoherence time $\sim$ O(10 $\mu s$) mentioned earlier in TMDCs at 5 -10 K [57-59] by a factor of $10^{-4}$-$10^{-5}$, allowing for successful error correction [72].

### V-4. Initialization, readout, and qugates

The SVO pseudospin qubit naturally shares properties of spin or valley qubits. As such, for initialization, readout, and qugate implementation one may adapt the methods previously developed for spin or valley qubits. For example, one may initialize the qubit by placing a "pseudospin valve" – the analogy to a spin valve in close proximity [14]. For readout, the spin-to-charge conversion scheme [73,74] could be adapted here as well. Last, in order to implement a two-qubit gate (qugate), one could place two qubits side by side, and make use of the electrically-tunable exchange coupling $J$ between localized pseudospins to perform a $\sqrt{SWAP}$ operation [14,15]. Overall, all-electric, universal SVO-based quantum computing is therefore feasible according to Divincenzo's criteria [72].

Last, we note that SVO qubits and qugates envisioned here can be realized with gated structures. This makes the corresponding quantum computing scalable. Combined with the optimized electrical manipulation time ~ O($ns$) in the case of in-plane configuration and the experimentally observed, much longer SVO decoherence time, favorable characteristics are implied for SVO-based quantum computing.

## VI. SUMMARY

In summary, for an insightful understanding and applications in spin-valley-orbital pseudospin physics, this work has formulated an effective theory, with important field effects included. Based on the theory, the linear response of a SVO pseudospin such as Zeeman and Rashba type effects has been discussed, with a clear connection established among the underlying band structure, external fields, and pseudospin physics.

Specifically, the work has investigated the pseudospin-flip coupling for pseudospin control, based on bare models that elucidate quantum paths leading to the coupling. Reduction of bare models yields the effective theory as well as expressions of effective coupling parameters in terms of band structure and bare coupling parameters. Two configurations, one with vertical and the other in-plane fields, are identified as of particular interest for pseudospin manipulation. The manipulation is shown, in the context of SVO-based quantum computing, to be achievable via electrical interaction mechanisms - SOI or VOI, and magnetic Zeeman effects. Overall, an optimized electrical manipulation time ~ O($ns$) is given.

In conclusion, field-modulatable spin-valley-orbital physics carries numerous promises. Together with the distinct electron-based spin-valley physics in the same material, it brings the rather appealing prospect - versatile spintronic type applications in a single material with flexible principles as well as carrier species.

## ACKNOWLEDGMENT

We would like to acknowledge the financial support of MoST, ROC through the Contract No. 107-2112-M-007 -023.

## APPENDIX A

## MATRIX ELEMENTS

The information of state symmetry as given in **Figure 4** and **Table 1** helps the evaluation of matrix elements. Below we provide examples of matrix elements used in our work and evaluated with this information:

(1) Matrix elements involving the spin operator, such as

$$\langle \Psi^n_{D_2(D_{-2})}, \downarrow, \tau | s_x | \Psi^m_{D_2(D_{-2})}, \uparrow, \tau \rangle = \delta_{nm}(s_x)_{\downarrow\uparrow}, \quad (A1)$$

(2) Matrix elements involving the momentum operator, such as

$$\langle \Psi^{(n)}_{D_{-2}}, s_z, K | p_+ | \Psi^{(m)}_{D_2}, s_z, K \rangle = 0,$$
$$\langle \Psi^{(n)}_{D_{-2}}, s_z, K | p_- | \Psi^{(m)}_{D_2}, s_z, K \rangle = 2P_{nm},$$
(A2)

$$\langle \Psi^{(i)}_{D_0}, s_z, K | p_+ | \Psi^{(j)}_{D_2}, s_z, K \rangle = 2P_{ij},$$
$$\langle \Psi^{(i)}_{D_0}, s_z, K | p_- | \Psi^{(j)}_{D_2}, s_z, K \rangle = 0,$$
(A3)

where the momentum matrix elements $P_{nm}$ and $P_{ij}$ are both imaginary numbers. In addition, we have

$$\langle \Psi^{(n)}_{D_{-2}}, s_z, K | p_{\pm} | \Psi^{(m)}_{D_2}, s_z, K \rangle$$
$$= -\langle \Psi^{(m)}_{D_{-2}}, -s_z, -K | p_{\pm} | \Psi^{(n)}_{D_2}, -s_z, -K \rangle,$$
(A4)

$$\langle \Psi^{(n)}_{D_0}, s_z, K | p_{\pm} | \Psi^{(m)}_{D_2}, s_z, K \rangle$$
$$= -\langle \Psi^{(m)}_{D_{-2}}, -s_z, -K | p_{\pm} | \Psi^{(n)}_{D_0}, -s_z, -K \rangle.$$
(A5)

(3) Matrix elements involving $z$, such as

$$\langle \Psi^{(n)}_{D_{\pm 2}(D_0)}, s_z, K | z | \Psi^{(m)}_{D_{\mp 1}(P_0)}, s_z, K \rangle = \zeta_{nm}, \quad (A6a)$$

$$\langle \Psi^{(n)}_{D_{-2}}, s_z, K | z | \Psi^{(m)}_{D_2}, s_z, K \rangle = 0, \quad (A6b)$$

where $\zeta_{nm}$ is a real number of O($a$). For time-reversal conjugated states, we have

$$\langle \Psi^{(n)}_{D_{\pm 2}}, s_z, K | z | \Psi^{(m)}_{D_{\mp 1}}, s_z, K \rangle$$
$$= -\langle \Psi^{(m)}_{D_{\mp 1}}, -s_z, -K | z | \Psi^{(n)}_{E_{\pm 2}}, -s_z, -K \rangle.$$
(A7a)

$$\langle \Psi^{(n)}_{D_0}, s_z, K | z | \Psi^{(m)}_{P_0}, s_z, K \rangle$$
$$= \langle \Psi^{(m)}_{P_0}, -s_z, -K | z | \Psi^{(n)}_{D_0}, -s_z, -K \rangle.$$
(A7b)

(4) Matrix elements involving the SOI, such as

$$\left\langle \Psi^{(n)}_{D_{-1}(D_1)}, \uparrow(\downarrow), \tau \left| \frac{\Delta^{(n,m)}_{so}}{4} \vec{L} \cdot \vec{s} \right| \Psi^{(m)}_{D_0(D_0)}, \downarrow(\uparrow), \tau \right\rangle = \sqrt{\frac{3}{8}} \Delta^{(n,m)}_{so},$$

(A8)

where the SOI parameter $\Delta_{so}^{(n,m)}$ is a real number. Moreover,

$$\left\langle \Psi_{D_{-1}(D_1)}^{(n)}, \uparrow(\downarrow), \tau \middle| \vec{L} \cdot \vec{s} \middle| \Psi_{D_0(D_0)}^{(m)}, \downarrow(\uparrow), \tau \right\rangle = \left\langle \Psi_{D_0(D_0)}^{(m)}, \uparrow(\downarrow), -\tau \middle| \vec{L} \cdot \vec{s} \middle| \Psi_{D_1(D_{-1})}^{(n)}, \downarrow(\uparrow), -\tau \right\rangle. \quad (A9)$$

## APPENDIX B

## TWELVE-STATE, EIGHT-STATE, AND EXTENDED BARE MODELS

We present firstly the minimal models that generate primary Hamiltonian terms in the effective theory. Following it, extension of the models for generating secondary terms is briefly discussed.

**Twelve-state model for the vertical configuration**

The Hamiltonian has been given by Eqn. (29) and is repeated below

$$H^{\perp} = H_{band} + H_{\vec{k}\cdot\vec{p}} + H_{SOI,off-diag} + H_{\varepsilon}^{\perp} + H_{Z}^{\perp} + U_{diag} + U_{K\leftrightarrow K'}. \quad (B1)$$

$H^{\perp}$ is the Hamiltonian. $H_{band}$ describes the "bare" energy bands of an electron and consists of only diagonal matrix elements given by

$$(H_{band})_{ij} = \left(E_i + p^2/2m_i^{\perp}\right)\delta_{ij} \quad (B2)$$

where $m_i^{\perp}$ is the "bare" mass and $E_i$ is the band edge energy for basis state $|\varphi_i\rangle$, including SOI-induced energy shift $\Delta_{so}^{(n_i,n_i)}$ ($n_i$ = band index of $|\varphi_i\rangle$). For the valence band, we write $\Delta_{so}^{(v,v)} = \Delta_{so}$, which is the spin-orbit gap, with $\Delta_{so} \sim$ 0.18 $eV$ in MoSe$_2$ and $\Delta_{so} \sim$ 0.46 $eV$ in WSe$_2$ [75]. $H_{\varepsilon}^{\perp}$ describes parity-mixing between states due to $\varepsilon_z$, with

$$(H_{\varepsilon}^{\perp})_{ij} = e\varepsilon_z \zeta_{ij},$$
$$\zeta_{i,j} = \langle\varphi_i|z|\varphi_j\rangle. \quad (B3)$$

$H_Z^{\perp}$ describes the Zeeman interaction due to $B_z$, with

$$(H_Z^{\perp})_{ij} = (1/2)g_i^{\perp}\mu_B B_z \delta_{ij}. \quad (B4)$$

$\mu_B$ is the Bohr magneton and $g_i^{\perp}$ is the "bare" g-factor. For reference, below we provide the explicit matrix form of $(H_{\vec{k}\cdot\vec{p}} + H_{SOI,off-diag} + H_{\varepsilon}^{\perp} + H_Z^{\perp})_{i\leq 6, j\leq 6}$

$$\begin{pmatrix} \frac{1}{2}g_1^{\perp}\mu_B B_z & \frac{\hbar}{m_e}k_+ P_{vn_1} & 0 & e\varepsilon_z\zeta_{vn_2} & 0 & 0 \\ \frac{\hbar}{m_e}k_- P_{vn_1}^* & \frac{1}{2}g_2^{\perp}\mu_B B_z & 0 & 0 & \sqrt{\frac{3}{8}}\Delta_{so}^{(n_1,n_3)} & 0 \\ 0 & 0 & \frac{1}{2}g_3^{\perp}\mu_B B_z & \sqrt{\frac{3}{8}}\Delta_{so}^{(n_1',n_2)} & 0 & \frac{\hbar}{m_e}k_+ P_{n_1'n_4} \\ e\varepsilon_z\zeta_{vn_2} & 0 & \sqrt{\frac{3}{8}}\Delta_{so}^{(n_1',n_2)} & \frac{1}{2}g_4^{\perp}\mu_B B_z & 0 & 0 \\ 0 & \sqrt{\frac{3}{8}}\Delta_{so}^{(n_1,n_3)} & 0 & 0 & \frac{1}{2}g_5^{\perp}\mu_B B_z & e\varepsilon_z\zeta_{n_3n_4} \\ 0 & 0 & \frac{\hbar}{m_e}k_- P_{n_1'n_4}^* & 0 & e\varepsilon_z\zeta_{n_3n_4} & \frac{1}{2}g_6^{\perp}\mu_B B_z \end{pmatrix}$$

(B5)

Here, $P_{vn_1}$ and etc. are the momentum matrix element parameters. In the presence of $B_z$, we make the minimal substitution $\hbar\vec{k} \to \vec{\Pi} = \vec{p} + e\vec{A}$ in the above matrix, where $\vec{A}$ is the corresponding vector potential. $\Delta_{so}^{(n_1,n_3)}$ and etc. are SOI strength parameters. Note that

$$(H_{\vec{k}\cdot\vec{p}})_{i\geq 7, j\geq 7} = -\left[(H_{\vec{k}\cdot\vec{p}})_{j-6,i-6}\right]^*,$$

$$(H_{SOI,off-diag} + H_Z^{\perp})_{i\geq 7, j\geq 7}$$
$$= (H_{SOI,off-diag} - H_Z^{\perp})_{j-6,i-6},$$

$$(H_{\varepsilon}^{\perp})_{i\geq 7, j\geq 7} = -\left[(H_{\varepsilon}^{\perp})_{j-6,i-6}\right]^*, \quad (B6)$$

due to the time reversal symmetry.

For parameters in $H^{\perp}$, the "bare" mass $m_1^{\perp}$ and g-factor $g_1^{\perp}$ are chosen in such a way to ensure that they are restored to "renormalized" valence band parameters $\{m^*, g_{eff}^{\perp}\}$, when the bare model is reduced to the effective theory for valence band. See **Appendix C**.

**Eight-state model for the in-plane configuration**

The Hamiltonian has been given by Eqn. (37) and is repeated below

$$H^{//} = H_{band} + H_{\vec{k}\cdot\vec{p}} + H_{SOI,off-diag} + U_{diag} + U_{K\leftrightarrow K'} + H_{VOI}^{//} + H_{\varepsilon}^{//} + H_Z^{//}. \quad (B7)$$

The first five terms on the right hand side have the same interpretations as their corresponding parts in $H^{\perp}$. $H_{VOI}^{//}$ is the VOI due to $\vec{\varepsilon}_{//}$ ($=(\varepsilon_x, \varepsilon_y)$), with

$$(H_{VOI}^{//})_{ij} = \delta_{ij}R_{VOI}^{(i)}\left(\varepsilon_y p_x - \varepsilon_x p_y\right),$$
$$R_{VOI}^{(i)} = -R_{VOI}^{(i+4)}, \quad (B8)$$

where $R_{VOI}^{(i)}$ is the "bare" coupling parameter. $H_\varepsilon^{//}$ is the electric potential energy due to $\vec{\varepsilon}_{//}$, with

$$(H_\varepsilon^{//})_{ij} = e\vec{\varepsilon}_{//} \cdot \vec{r}_{//} \delta_{ij}. \tag{B9}$$

$H_Z^{//}$ describes the Zeeman interaction due to $\vec{B}_{//}$ ($=(B_x, B_y)$), with

$$(H_Z^{//})_{ij} = (1/2) g_e \mu_B \left[ (s_x)_{ij} B_x + (s_y)_{ij} B_y \right]. \tag{B10}$$

We provide $(H_{\vec{k}\cdot\vec{p}} + H_{SOI,off-diag} + H_Z^{//} + H_{VOI}^{//})_{i\leq 4, j\leq 4}$ explicitly below:

$$\begin{pmatrix} R_{VOI}^{(1)} |\vec{\varepsilon}_{//}| \left( n_y^{(\varepsilon)} p_x - n_x^{(\varepsilon)} p_y \right) & \frac{1}{2} g_e \mu_B |\vec{B}_{//}| n_-^{(B)} & \frac{\hbar}{m_e} k_- P_{vn} & 0 \\ \frac{1}{2} g_e \mu_B |\vec{B}_{//}| n_+^{(B)} & R_{VOI}^{(2)} |\vec{\varepsilon}_{//}| \left( n_y^{(\varepsilon)} p_x - n_x^{(\varepsilon)} p_y \right) & 0 & \frac{\hbar}{m_e} k_- P_{vn} \\ \frac{\hbar}{m_e} k_+ P_{vn}^* & 0 & R_{VOI}^{(3)} |\vec{\varepsilon}_{//}| \left( n_y^{(\varepsilon)} p_x - n_x^{(\varepsilon)} p_y \right) & \frac{1}{2} g_e \mu_B |\vec{B}_{//}| n_-^{(B)} \\ 0 & \frac{\hbar}{m_e} k_+ P_{vn}^* & \frac{1}{2} g_e \mu_B |\vec{B}_{//}| n_+^{(B)} & R_{VOI}^{(4)} |\vec{\varepsilon}_{//}| \left( n_y^{(\varepsilon)} p_x - n_x^{(\varepsilon)} p_y \right) \end{pmatrix}$$

(B11)

Here, $(n_x^{(\varepsilon)}, n_x^{(\varepsilon)}) = (\varepsilon_x / |\vec{\varepsilon}_{//}|, \varepsilon_y / |\vec{\varepsilon}_{//}|)$, $n_\pm^{(B)} = n_x^{(B)} \pm i n_y^{(B)}$, and $(n_x^{(B)}, n_x^{(B)}) = (B_x / |\vec{B}_{//}|, B_y / |\vec{B}_{//}|)$. Note that

$$(H_{\vec{k}\cdot\vec{p}})_{i\geq 5, j\geq 5} = -\left[ (H_{\vec{k}\cdot\vec{p}})_{j-4, i-4} \right]^*,$$

$$(H_Z^{//})_{i\geq 5, j\geq 5} = \left[ (H_Z^{//})_{j-4, i-4} \right]^*,$$

$$(H_{SOI,off-diag} + H_{VOI}^{//})_{i\geq 5, j\geq 5} = (H_{SOI,off-diag} - H_{VOI}^{//})_{j-4, i-4},$$

(B12)

due to the time reversal symmetry. In the choice of various parameters in $H^{//}$, the "bare" $\{m_1^{//}, g_e, R_{VOI}^{(1)}\}$ are chosen to give "renormalized" $\{m^*, g_{eff}^{(3)//}, R_{VOI,eff}\}$ when the model is reduced to the effective theory for valence band. See **Appendix C**.

**Extended models**

Primary Hamiltonian terms in $H_{eff}^{(off-diag)}$ are $\{g_{eff}^{(2)//}, g_{eff}^{(3)//}, R_{SOI,eff}^{(4)\perp}\}$-dependent. At the second order of perturbation theory, it can be verified that only **Class B** paths contribute to the $g_{eff}^{(2)//}$-dependent term, which are the Type-IV paths shown in **Figure 7**. In contrast, at the third and fourth orders, **Class A** paths shown in **Figures 6** and **7** contribute to $\{g_{eff}^{(3)//}, R_{SOI,eff}^{(4)\perp}\}$-dependent terms.

Apart from the **Class A** paths in **Figures 6 and 7**, it can be verified that **Class B** paths exist at the third and fourth orders. These additional paths lead to $\{g_{eff}^{(3,corr)//}, R_{SOI,eff}^{(4,corr)\perp}\}$-dependent terms. Such paths use intermediate states outside those already included in the minimal models. Therefore, the extension of models for deriving secondary terms consists of identifying these additional paths and states, and adding the states to basis state sets of the models.

In some cases, a conjugated relation exists based on which **Class B** paths can be built from **Class A** ones in a systematic way. Consider the following **Class A** path: $(VBM, K) \xrightarrow{s_x} (\Psi_{D_2}^{(v)}, \downarrow, K) \xrightarrow{\vec{k}\cdot\vec{p}=\frac{1}{2}(k_+ p_- + k_- p_+)} (\Psi_{D_{-2}}^{(n)}, \downarrow, K) \xrightarrow{U_{elastic}} (VBM, K')$, which is shown in **Figure 7** as a Type-II path for the $g_{eff}^{(3)//}$-dependent term. In this path, the $\vec{k}\cdot\vec{p}$ coupling beween the states of $(D_2, \downarrow, K)$ and $(D_{-2}, \downarrow, K)$ comes from the "$k_-p_+$" term, as can be verified using **Appendix A**. However, through the alternative "$k_+p_-$" term, it can instead connect the $(D_2, \downarrow, K)$ state to an $(D_0, \downarrow, K)$ state, and then arrive at $(VBM, K')$ via the $U_{elastic}$-induced valley-flip scattering. With $(VBM, K')$ a state of $D_{-2}$, this generates an alternative path – a **Class B** one where the irreducible representation index is varied from $D_0$ to $D_{-2}$ during the valley flip, and thus contributes to the $g_{eff}^{(3,corr)//}$-dependent term. To account for this alternative path, the new intermediate state of $(D_0, \downarrow, K)$ would have to be added to the bare model. With an analysis such as the above and beyond, we identify all **Class B** paths that contribute to secondary terms, and expand the basis state sets to those of twenty-four and twelve states, for the vertical and in-plane configurations, respectively. The additional states are given by

$$\left\{ \left| \Psi_{D_1}^{(n_2')}, \uparrow, K \right\rangle, \left| \Psi_{D_{-1}}^{(n_3')}, \downarrow, K \right\rangle, \left| \Psi_{D_{-2}}^{(n_4')}, \uparrow, K \right\rangle, \left| \Psi_{D_2}^{(n_5)}, \downarrow, K \right\rangle, \left| \Psi_{P_0}^{(n_6)}, \uparrow, K \right\rangle, \right.$$
$$\left| \Psi_{P_0}^{(n_6')}, \downarrow, K \right\rangle \left| \Psi_{D_{-1}}^{(n_2')}, \downarrow, K' \right\rangle, \left| \Psi_{D_1}^{(n_3')}, \uparrow, K' \right\rangle, \left| \Psi_{D_2}^{(n_4')}, \downarrow, K' \right\rangle, \left| \Psi_{D_{-2}}^{(n_5)}, \uparrow, K' \right\rangle,$$
$$\left. \left| \Psi_{P_0}^{(n_6)}, \downarrow, K' \right\rangle, \left| \Psi_{P_0}^{(n_6')}, \uparrow, K' \right\rangle \right\},$$

(B13)

in the vertical case; and

$$\left\{ \left| \Psi_{D_0}^{(n_1)}, \uparrow, K \right\rangle, \left| \Psi_{D_0}^{(n_1)}, \downarrow, K \right\rangle, \left| \Psi_{D_0}^{(n_1)}, \downarrow, K' \right\rangle, \left| \Psi_{D_0}^{(n_1)}, \uparrow, K' \right\rangle \right\},$$

(B14)

in the in-plane case.

Last, we note that the bare Hamiltonian operators in the extension remain the same forms as those in minimal models and so will not be redundantly presented.

## APPENDIX C

### THE SCHRIEFFER-WOLFF REDUCTION

The Schrieffer-Wolff (SW) reduction provides a way to obtain from a bare model the effective Hamiltonian in a reduced subspace. [76] In **C-1**, we summarize the SW reduction in general. In **C-2**, we apply the method to the case of ac field-driven qubits, which was discussed in **V-1** and **V-2** of the main text in the adiabatic approach, for a verification of the approach. In **C-3**, the method is applied to the derivation of effective coupling parameters.

**C-1. General result**

We consider a general Hamiltonian in the perturbation theory,

$$H = H_0 + H_1 + X_1, \qquad (C1)$$

where $H_0$ describes the unperturbed system, with eigenstates $\{|m\rangle\text{'s}\}$ and eigenvalues $\{E_m\text{'s}\}$. $\{|m\rangle\text{'s}\}$ are used below as basis functions. $H_1$ is a time-independent perturbation, and $X_1$ is some additional perturbation of interest which could be time-dependent or independent. We take the diagonal $(H_1)_{nn} = 0$ for simplicity.

Denote the subspace of interest with A, which is spanned by $\{|n\rangle\text{'s}, n=1,...,\alpha\}$, and the subspace complementary to A with B. The SW reduction consists of performing a similarity transformation on $H$, yielding the effective Hamiltonian

$$\begin{aligned}H_{eff}^{(A)} &= e^S[H_0 + H_1 + X_1]e^{-S}\\ &= H_0 + H_1 + X_1\\ &\quad + (1/2)[S,H_1] + [S,X_1]\end{aligned} \qquad (C2a)$$

$$+\frac{1}{2}\{S^2, X_1\} + \frac{1}{6}[S^3, X_1] - SX_1S - \frac{1}{2}[S, SX_1S] \qquad (C2b)$$

$+......$

(C2)

in the subspace of A, where

$$S = S_1 + S_2 + S_3 + ...,$$
$$(S_1)_{nm} = \frac{(H_1)_{nm}}{E_n - E_m} = O(H_1),$$
$$(S_2)_{nm} = O(H_1^2),$$
$$(S_3)_{nm} = O(H_1^3),$$

(C3)

with $|n\rangle$ and $|m\rangle$ above belonging to A and B, respectively. $S_1$, $S_2$, and $S_3$ are, respectively, of $O(H_1)$, $O(H_1^2)$, and $O(H_1^3)$, and used to remove the $H_1$-induced coupling between A and B up to $O(H_1)$, $O(H_1^2)$, and $O(H_1^3)$, respectively. For complete expressions of $S_2$ and $S_3$, see Reference 77. Below, we provide only partial expressions

$$(S_2)_{nm} = \frac{(H_1)_{nm'}(H_1)_{m'm}}{(E_n - E_m)(E_n - E_{m'})} + ......,$$

$$(S_3)_{nm} = \frac{(H_1)_{nm'}(H_1)_{m'm''}(H_1)_{m''m}}{(E_n - E_m)(E_n - E_{m'})(E_n - E_{m''})} + ......$$

(C4)

which actually enter our study. The discussions in **C-2** and **C-3** are based on Eqns. (C1)-(C4).

Eqns. (C2a) and (C2b) have important implications for this work. For example, due to the presence of $[S, X_1]$ ($[S, H_1]$) in $H_{eff}^{(A)}$, it shows that when $X_1 = 0$ ($H_1 = 0$) in the subspace of A, $[S, X_1]$ ($[S, H_1]$) could still provide an "effective coupling" of $O(H_1)O(X_1)$ ($O(H_1)^2$) between states in A. Generalization to effective couplings of $O(H_1^2)O(X_1)$ and $O(H_1^3)O(X_1)$ can be obtained from Eqn. (C2b) and will be used in **C-3** for discussions there.

**C-2. AC-field driven qubits**

Consider now the QD envisioned in our work, which is subject to the potential energy $U_{quad}(x) + U_{QW}(y) + e\varepsilon_{ac}x\cos(\omega_{ac}t)$. For reference, we reproduce Eqn. (37) below

$$\begin{aligned}&U_{quad}(x) + e\varepsilon_{ac}x\cos(\omega_{ac}t)\\ &= (1/2)m^*\omega_x^2[x - x_0(t)]^2 + O(\varepsilon_{ac})^2, \qquad (37)\\ &x_0(t) = -e\varepsilon_{ac}\cos(\omega_{ac}t)/m^*\omega_x^2.\end{aligned}$$

Correct up to $O(\varepsilon_{ac})$, the equation describes a QD that oscillates at the frequency $\omega_{ac}$. It suggests us to work with the transformed coordinates, namely, $x' = x - x_0(t)$, $y' = y$, and $t' = t$, in a reference frame moving synchronically with the QD. Denote $H_{x,y,t}$ as the QD Hamiltonian in the lab reference frame, with the corresponding Hamiltonian equation $H_{x,y,t}\psi = i\hbar\partial_t\psi$. Then, in the moving reference frame, it transforms to

$$\begin{aligned}H_{x',y',t'}\psi &= i\hbar\partial_{t'}\psi,\\ H_{x',y',t'} &= H_{x,y,t}|_{x\to x'+x_0(t), y\to y', t\to t'} - p_x\partial_t x_0|_{t\to t'}.\end{aligned}$$
(C5)

For simplicity, we switch the notation $(x', y', t')$ back to $(x, y, t)$. Then, overall, correct up to $O(\varepsilon_{ac})$, the transformation replaces the ac potential energy "$e\varepsilon_{ac}x\cos(\omega_{ac}t)$" by "$-p_x\partial_t x_0$" in the Hamiltonian. Below, we apply the result of Eqn. (C5).

In the vertical-field case, the QD ground states $\{|K, m=0, n=0\rangle, |K', m=0, n=0\rangle\}$ ($= \{|K\rangle_{QD}, |K'\rangle_{QD}\}$) are used as qubit basis states. In order to obtain the effective Hamiltonian in the qubit state subspace, we take $H_0 = H_{eff}^{(diag)}$, $H_1 = H_{eff}^{(off-diag)}$, and $X_1 = -p_x\partial_t x_0$, and perform the SW reduction with $S = S_1$. It leads to the following effective coupling

$$\begin{aligned}&\langle K, 0, 0|[S_1, X_1]|K', 0, 0\rangle\\ &= \frac{\langle K, 0, 0|H_{eff}^{(off-diag)}|K', 1, 0\rangle\langle K', 1, 0|-p_x\partial_t x_0|K', 0, 0\rangle}{E_{Z,eff}^\perp - \hbar\omega_x}\\ &\quad - \frac{\langle K, 0, 0|-p_x\partial_t x_0|K, 1, 0\rangle\langle K, 1, 0|H_{eff}^{(off-diag)}|K', 0, 0\rangle}{E_{Z,eff}^\perp + \hbar\omega_x}\end{aligned}$$

$$= i\frac{m^*}{\hbar}\alpha^\perp\big|_{n=0}\frac{dx_0}{dt}.$$

(C6)

Above, we have used the identities in **Appendix A** to simplify the expression.

In the in-plane-field case, $\{|+, m=0, n=0\rangle, |-, m=0, n=0\rangle\}$ $(=\{|+\rangle_{QD}, |-\rangle_{QD}\})$ are used as qubit basis states. In order to obtain the qubit Hamiltonian, we take $H_0 = H_{eff}(\varepsilon_y = 0)$, $H_1 = H_{VOI}^{(val)} + e\varepsilon_y y$, and $X_1 = -p_x \partial_t x_0$, and perform the SW reduction with $S = S_1$. It leads to the following effective coupling

$$\langle +,0,0|[S_1, X_1]|-,0,0\rangle$$
$$= \frac{\langle +,0,0|H_{VOI}^{(val)}|-,1,0\rangle\langle -,1,0|-p_x\partial_t x_0|-,0,0\rangle}{-E_Z^{//} - \hbar\omega_x}$$
$$- \frac{\langle +,0,0|-p_x\partial_t x_0|+,1,0\rangle\langle +,1,0|H_{VOI}^{(val)}|-,0,0\rangle}{-E_Z^{//} + \hbar\omega_x}$$
$$= \frac{m^*}{\hbar}\alpha^{//}\frac{dx_0}{dt}.$$

(C7)

Eqns. (C6) and (C7) confirm Eqns. (48) and (56) obtained in the adiabatic approximation, respectively.

**C-3. Coupling parameters**

We take A (the subspace of interest) = {|VBM, K>, |VBM, K'>} below.

**Effective mass $m^*$**

In the $\vec{k}\cdot\vec{p}$ theory, an effective (or "re-normalized") mass consists of the "bare" mass and second-order corrections due to the perturbation "$\hbar\vec{k}\cdot\vec{p}/m_e$". Below, we provide for the valence band the relation between bare mass parameters "$\{m_1^\perp, m_1^{//}\}$" and effective mass $m^*$.

We apply the twelve-state model first. We take $\varepsilon_z = 0$, $B_z = 0$, $H_0 = H_{band}$, and the perturbation $H_1 = H_{\vec{k}\cdot\vec{p}}$. We perform the SW reduction with $S = S_1$, and obtain the valence band dispersion near K

$$E_v(k; \tau = K) = (H_0)_{11} + \frac{1}{2}([S_1, H_1])_{11}$$
$$= \frac{(\hbar k)^2}{2m_1^\perp} + \left(\frac{\hbar}{m_e}\right)^2 \sum_l \frac{\langle VBM, K|\vec{k}\cdot\vec{p}|\Psi_{D_0}^{(l)}, \uparrow, K\rangle\langle\Psi_{D_0}^{(l)}, \uparrow, K|\vec{k}\cdot\vec{p}|VBM, K\rangle}{(E_{VBM} - E_{l,\uparrow,K})},$$

(C8)

yielding

$$\frac{1}{m^*} = \frac{1}{m_1^\perp} + 2\left(\frac{1}{m_e}\right)^2 \sum_{n\in D_0} \frac{|P_{vn}|^2}{(E_{VBM} - E_{n,\uparrow,K})}.$$

(C9)

Above, we have summed over all intermediate states of $D_0$ representation for leading-order contributions. For reference, we provide $m_1^\perp$ below

$$\frac{1}{m_1^\perp} = \frac{1}{m_e} + 2\left(\frac{1}{m_e}\right)^2 \sum_{l\in D_{-2}} \frac{|P_{vl}|^2}{(E_{VBM} - E_{l,\uparrow,K})},$$

(C10)

without deriving it. The above discussion could also be performed using valley-K' states, which would yield identical results due to the time-reversal symmetry.

Similarly, one can work in the eight-state model and derive the relation between $m_1^{//}$ and $m^*$, namely,

$$\frac{1}{m^*} = \frac{1}{m_1^{//}} + 2\left(\frac{1}{m_e}\right)^2 \sum_{n\in D_{-2}} \frac{|P_{vn}|^2}{(E_{VBM} - E_{l,\uparrow,K})},$$
$$\frac{1}{m_1^{//}} = \frac{1}{m_e} + 2\left(\frac{1}{m_e}\right)^2 \sum_{l\in D_0} \frac{|P_{vl}|^2}{(E_{VBM} - E_{l,\uparrow,K})}.$$

(C11)

**g-factor in the vertical Zeeman effect**

We work in the twelve-state mode with the presence of $B_z$. We make the minimal substitution $\hbar\vec{k} \to \vec{\Pi} = \vec{p} + e\vec{A}$. Following the same procedure in deriving effective mass above, we obtain

$$E_v(\hbar\vec{k} \to \vec{\Pi}, \tau) = \frac{\vec{\Pi}^2}{2m^*} + \frac{1}{2}\tau g_{eff}^\perp \mu_B B_z,$$

$$g_{eff}^\perp = g_e + g_{valley-orbital}^\perp$$

$$= g_1^\perp - \frac{4}{m_e}\sum_{l\in D_0}\frac{|P_{vl}|^2}{(E_{VBM} - E_{l,\uparrow,K})},$$

$$g_1^\perp = g_e + \frac{4}{m_e}\sum_{l\in D_{-2}}\frac{|P_{vl}|^2}{(E_{VBM} - E_{l,\uparrow,K})},$$

$$g_{valley-orbital}^\perp = \frac{4}{m_e}\left[\sum_{l\in E_{-2}}\frac{|P_{vl}|^2}{(E_{VBM} - E_{l,\uparrow,K})} - \sum_{l\in D_0}\frac{|P_{vl}|^2}{(E_{VBM} - E_{l,\uparrow,K})}\right].$$

(C12)

**$R_{SOI,eff}^{(4)\perp}$ in the vertical Rashba effect**

We work in the twelve-state model, and take $B_z = 0$, $H_0 = H_{band}$, $H_1 = H_{\vec{k}\cdot\vec{p}} + H_{SOI,off-diag} + H_\varepsilon^\perp$, and $X_1 = U_{K\leftrightarrow K'}$. As the effect involves fourth-order quantum paths, we collect the terms of $O(H_1)^3 O(X_1)$ in Eqns. (C2a) and (C2b) and obtain

$$H_{eff}^{(off-diag)}$$
$$= [S_3, X_1] + \frac{1}{2}\{S_1 S_2 + S_2 S_1, X_1\} + \frac{1}{6}[S_1^3, X_1]\cdot$$
$$- S_2 X_1 S_1 - S_1 X_1 S_2 - \frac{1}{2}[S_1, S_1 X_1 S_1]$$

(C13)

We find

$$\left(H_{\text{eff}}^{(\text{off}-\text{diag})}\right)_{21}$$
$$= \sum_{\substack{n_1',n_2,n_4 \\ \text{for Path Type-I}}} \left[ \langle VBM, K'|U_{K\leftrightarrow K'}|\Psi_{D_{-2}}^{(n_4)},\downarrow,K\rangle \langle \Psi_{D_{-2}}^{(n_4)},\downarrow,K|\frac{\hbar}{m_e}\vec{k}\cdot\vec{p}|\Psi_{D_0}^{(n_1')},\downarrow,K\rangle \right.$$
$$\langle \Psi_{D_0}^{(n_1')},\downarrow,K|\frac{\Delta_{so}^{(n_1',n_2)}}{4}\vec{L}\cdot\vec{s}|\Psi_{D_{-1}}^{(n_2)},\uparrow,K\rangle \langle \Psi_{D_{-1}}^{(n_2)},\uparrow,K|e\varepsilon_z z|VBM,K\rangle$$
$$/(E_{VBM}-E_{n_2,\uparrow,K})(E_{VBM}-E_{n_1',\downarrow,K})(E_{VBM}-E_{n_4,\downarrow,K})$$
$$+\langle VBM,K'|e\varepsilon_z z|\Psi_{D_1}^{(n_2)},\downarrow,K'\rangle \langle \Psi_{D_1}^{(n_2)},\downarrow,K'|\frac{\Delta_{so}^{(n_2,n_1')}}{4}\vec{L}\cdot\vec{s}|\Psi_{D_0}^{(n_1')},\uparrow,K'\rangle$$
$$\langle \Psi_{D_0}^{(n_1')},\uparrow,K'|\frac{\hbar}{m_e}\vec{k}\cdot\vec{p}|\Psi_{D_2}^{(n_4)},\uparrow,K'\rangle \langle \Psi_{D_2}^{(n_4)},\uparrow,K'|U_{K\leftrightarrow K'}|VBM,K\rangle$$
$$\left. /(E_{VBM}-E_{n_1',\uparrow,K'})(E_{VBM}-E_{n_2,\downarrow,K'})(E_{VBM}-E_{n_4,\uparrow,K'})\right]$$
$$+[\text{similar contributions from other path types}],$$

(C14)

After simplification, we make the substitution $\hbar k_- \to \Pi_-$ for a finite $B_z$, and obtain in the linear regime

$$\left(H_{\text{eff}}^{(\text{off}-\text{diag})}\right)_{21} = \frac{i\varepsilon_z R_{SOI,\text{eff}}^{(4)\perp}}{\overline{\Delta}}\left\{U_{\text{elastic}}^{(IR-\text{diag})}e^{2iKy},\Pi_-\right\},$$

(C15)

where

$$R_{SOI,\text{eff}}^{(4)\perp} = \sqrt{\frac{3}{8}}\overline{\Delta}\left(\frac{-i}{m_e}\right)$$
$$\left\{\left[\sum_{\substack{n_1'\in D_0, \\ n_2\in D_{-1}, \\ n_4\in D_{-2}}} \frac{\lambda_{v,n_4}\Delta_{so}^{(n_1',n_2)}(e\zeta_{n_2 v})P_{n_4 n_1'}}{(E_{VBM}-E_{n_2,\uparrow,K})(E_{VBM}-E_{n_1',\downarrow,K})(E_{VBM}-E_{n_4,\downarrow,K})}\right]\right.$$
$$+\left[\sum_{\substack{n_1\in D_0, \\ n_3\in D_1, \\ n_4\in D_{-2}}} \frac{\lambda_{v,n_4}\Delta_{so}^{(n_3,n_1)}(e\zeta_{n_4 n_3})P_{n_1 v}}{(E_{VBM}-E_{n_1,\uparrow,K})(E_{VBM}-E_{n_3,\downarrow,K})(E_{VBM}-E_{n_4,\downarrow,K})}\right]$$
$$+\left[\sum_{\substack{n_1,n_1'\in D_0, \\ n_2\in D_{-1}}} \frac{\lambda_{n_1',n_1}\Delta_{so}^{(n_1',n_2)}(-e\zeta_{n_2 v})P_{n_1 v}}{(E_{VBM}-E_{n_1,\uparrow,K})(E_{VBM}-E_{n_2,\uparrow,K})(E_{VBM}-E_{n_1',\downarrow,K})}\right]$$
$$\left.+\left[\sum_{\substack{n_1\in D_0, \\ n_2\in D_{-1}, \\ n_3\in D_1}} \frac{\lambda_{n_2,n_3}\Delta_{so}^{(n_3,n_1)}(-e\zeta_{n_2 v})P_{n_1 v}}{(E_{VBM}-E_{n_2,\uparrow,K})(E_{VBM}-E_{n_3,\downarrow,K})(E_{VBM}-E_{n_1,\uparrow,K})}\right]\right\},$$

(C16)

It can be verified that $R_{SOI,\text{eff}}^{(4)\perp}$ is real, with

$$R_{SOI,\text{eff}}^{(4)\perp} = O(\frac{\Delta_{so}ea^2}{\hbar\overline{\Delta}}).$$

(C17)

Here, $\overline{\Delta}$ = a typical gap.

### $g_{\text{eff}}^{(2)//}$ in the in-plane Zeeman effect

We apply the eight-state model, and take $\vec{\varepsilon}_{//} = 0$, $H_0 = H_{band}$, $H_1 = H_Z^{//}$, and $X_1 = U_{K\leftrightarrow K'}$. As the effect involves second-order quantum paths, we collect the terms of $O(H_1)O(X_1)$ in Eqn. (C2a), and obtain

$$H_{\text{eff}}^{(\text{off}-\text{diag})} = [S_1, X_1].$$

(C18)

We find

$$\left(H_{\text{eff}}^{(\text{off}-\text{diag})}\right)_{21}$$
$$= \frac{\langle VBM,K'|U_{K\leftrightarrow K'}|\Psi_{D_2}^{(v)},\downarrow,K\rangle \langle \Psi_{D_2}^{(v)},\downarrow,K|\frac{1}{2}g_e\mu_B\vec{B}_{//}\cdot\vec{s}_{//}|VBM,K\rangle}{(E_{VBM}-E_{v,\downarrow,K})}$$
$$+\frac{\langle VBM,K'|\frac{1}{2}g_e\mu_B\vec{B}_{//}\cdot\vec{s}_{//}|\Psi_{D_{-2}}^{(v)},\uparrow,K'\rangle \langle \Psi_{D_{-2}}^{(v)},\uparrow,K'|U_{K\leftrightarrow K'}|VBM,K\rangle}{(E_{VBM}-E_{v,\uparrow,K'})},$$

(C19)

After simplification, we finally obtain

$$\left(H_{\text{eff}}^{(\text{off}-\text{diag})}\right)_{21} = \frac{a}{\overline{\Delta}}g_{\text{eff}}^{(2)//}\mu_B B_+\left(U_{\text{elastic}}^{(IR-\text{flip})}\right)_+ e^{2iKy},$$

(C20)

where

$$g_{\text{eff}}^{(2)//} = \frac{\overline{\Delta}\lambda_{vv}g_e}{(E_{VBM}-E_{v,\downarrow,K})}.$$

(C21)

It can be verified that $g_{\text{eff}}^{(2)//}$ is real, with

$$g_{\text{eff}}^{(2)//} = O\left[\frac{\overline{\Delta}}{\Delta_{so}}\overline{\lambda}^{(IR-\text{flip})}\right]g_e,$$

(C22)

where $\overline{\lambda}^{(IR-\text{flip})}$ is the typical value of $\lambda_{n,l}$, the coupling strength for valley-flip scattering between bands of indices $n$ and $l$, in the case where the irreducible representation index of electron state varies in the scattering. See **Appendix D**.

### $g_{\text{eff}}^{(3)//}$ in the in-plane Zeeman effect

We apply the eight-state model, and take $\vec{\varepsilon}_{//} = 0$, $H_0 = H_{band}$, $H_1 = H_{\vec{k}\cdot\vec{p}} + H_Z^{//}$, and $X_1 = U_{K\leftrightarrow K'}$. As the effect involves third-order quantum paths, we collect the terms of $O(H_1)^2 O(X_1)$ in Eqns. (C2a) and (C2b), and obtain

$$H_{\text{eff}}^{(\text{off-diag})} = [S_2, X_1] + \frac{1}{2}\{S_1^2, X_1\} - S_1 X_1 S_1. \tag{C23}$$

We find

$$\begin{aligned}
&\left(H_{\text{eff}}^{(\text{off-diag})}\right)_{21} \\
&= \sum_{n \text{ for Path Type-I}} \Big[ \langle VBM, K'|U_{K\leftrightarrow K'}|\Psi_{D_{-2}}^{(n)}, \downarrow, K\rangle \langle \Psi_{D_{-2}}^{(n)}, \downarrow, K|\tfrac{1}{2}g_e\mu_B \vec{B}_{//}\cdot\vec{s}_{//}|\Psi_{D_{-2}}^{(n)}, \uparrow, K\rangle \\
&\quad \langle \Psi_{D_{-2}}^{(n)}, \uparrow, K|\tfrac{\hbar}{m_e}\vec{k}\cdot\vec{p}|VBM, K\rangle /(E_{VBM}-E_{n,\uparrow,K})(E_{VBM}-E_{n,\downarrow,K}) \\
&\quad + \langle VBM, K'|\tfrac{\hbar}{m_e}\vec{k}\cdot\vec{p}|\Psi_{D_{-2}}^{(n)}, \downarrow, K'\rangle \langle \Psi_{D_{-2}}^{(n)}, \downarrow, K'|\tfrac{1}{2}g_e\mu_B \vec{B}_{//}\cdot\vec{s}_{//}|\Psi_{D_{-2}}^{(n)}, \uparrow, K'\rangle \\
&\quad \langle \Psi_{D_{-2}}^{(n)}, \uparrow, K'|U_{K\leftrightarrow K'}|VBM, K\rangle /(E_{VBM}-E_{n,\downarrow,K'})(E_{VBM}-E_{n,\uparrow,K'}) \Big] \\
&+ [\text{similar contributions from other path types}],
\end{aligned} \tag{C24}$$

After simplification, we finally obtain

$$\left(H_{\text{eff}}^{(\text{off-diag})}\right)_{21} = i\frac{a}{\hbar\overline{\Delta}} g_{\text{eff}}^{(3)//} \mu_B B_+ \left[U_{\text{elastic}}^{(IR-diag)} e^{2iKy}, p_+\right], \tag{C25}$$

where

$$\begin{aligned}
g_{\text{eff}}^{(3)//} = -i\frac{g_e \hbar \overline{\Delta}}{2m_e a} \sum_{n\in D_{-2}} \lambda_{v,n} P_{nv} \Bigg[ &\frac{1}{(E_{VBM}-E_{n,\uparrow,K})(E_{VBM}-E_{n,\downarrow,K})} \\
&+ \frac{1}{(E_{VBM}-E_{n,\downarrow,K})(E_{VBM}-E_{v,\downarrow,K})} \\
&+ \frac{1}{(E_{VBM}-E_{v,\downarrow,K})(E_{VBM}-E_{n,\uparrow,K})} \Bigg].
\end{aligned} \tag{C26}$$

It can be verified that $g_{\text{eff}}^{(3)//}$ is real, with

$$g_{\text{eff}}^{(3)//} = O\left[\frac{\overline{\Delta}}{\Delta_{so}}\right] g_e. \tag{C27}$$

### $R_{VOI,\text{eff}}$ in the in-plane Rashba effect

We apply the eight-state model, and take $\vec{B}_{//} = 0$, $H_0 = H_{\text{band}}$, $H_1 = H_{\vec{k}\cdot\vec{p}}$, and $X_1 = e\vec{\varepsilon}_{//}\cdot\vec{r}$. This effect basically involves third-order quantum paths. We collect the terms of $O(H_1)^2 O(X_1)$ in Eqns. (C2a) and (C2b), and obtain

$$\left(H_{VOI}^{(val)}\right)_{11} - R_{VOI}^{(1)}\left(\vec{p}\times\vec{\varepsilon}_{//}\right)_z = [S_2, X_1] + \frac{1}{2}\{S_1^2, X_1\} - S_1 X_1 S_1$$

$$\begin{aligned}
&= \sum_{n\in D_{-2}} \Bigg\{ \left(\frac{-1}{2}\right)\Big[\langle VBM, K|e\vec{\varepsilon}_{//}\cdot\vec{r}|VBM, K\rangle \langle VBM, K|\tfrac{\hbar}{m_e}\vec{k}\cdot\vec{p}|\Psi_{D_{-2}}^{(n)}, \uparrow, K\rangle \\
&\quad \langle \Psi_{D_{-2}}^{(n)}, \uparrow, K|\tfrac{\hbar}{m_e}\vec{k}\cdot\vec{p}|VBM, K\rangle \\
&\quad + \langle VBM, K|\tfrac{\hbar}{m_e}\vec{k}\cdot\vec{p}|\Psi_{D_{-2}}^{(n)}, \uparrow, K\rangle \langle \Psi_{D_{-2}}^{(n)}, \uparrow, K|\tfrac{\hbar}{m_e}\vec{k}\cdot\vec{p}|VBM, K\rangle \\
&\quad \langle VBM, K|e\vec{\varepsilon}_{//}\cdot\vec{r}|VBM, K\rangle \Big] \\
&\quad + \langle VBM, K|\tfrac{\hbar}{m_e}\vec{k}\cdot\vec{p}|\Psi_{D_{-2}}^{(n)}, \uparrow, K\rangle \langle \Psi_{D_{-2}}^{(n)}, \uparrow, K|e\vec{\varepsilon}_{//}\cdot\vec{r}|\Psi_{D_{-2}}^{(n)}, \uparrow, K\rangle \\
&\quad \langle \Psi_{D_{-2}}^{(n)}, \uparrow, K|\tfrac{\hbar}{m_e}\vec{k}\cdot\vec{p}|VBM, K\rangle \Bigg\}/(E_{VBM}-E_{n,\uparrow,K})^2
\end{aligned} \tag{C28}$$

After simplification, we finally obtain

$$\left(H_{VOI}^{(val)}\right)_{11} = R_{VOI,\text{eff}}\left(\vec{p}\times\vec{\varepsilon}_{//}\right)_z, \tag{C29}$$

where

$$R_{VOI,\text{eff}} = R_{VOI}^{(1)} - \frac{e\hbar}{2m_e^2} \sum_{n\in D_{-2}} \frac{|P_{vn}|^2}{(E_{VBM}-E_{n,\uparrow,K})^2} \tag{C30}$$

For reference, we provide

$$R_{VOI}^{(1)} = \frac{e\hbar}{2m_e^2} \sum_{l\in D_0} \frac{|P_{vl}|^2}{(E_{VBM}-E_{l,\uparrow,K})^2} \tag{C31}$$

without deriving it. It can be verified that $R_{VOI,\text{eff}}$ is real, and estimated to be

$$R_{VOI,\text{eff}} = O\left[\frac{e\hbar}{m_e^2}\frac{|P_{vc}|^2}{\overline{\Delta}^2}\right]. \tag{C32}$$

In order to obtain secondary parameters, one performs the SW reduction on extended bare models. We provide results below without giving details.

### $R_{SOI,\text{eff}}^{(4,\text{corr})\perp}$ in the vertical Rashba effect

Complete expressions are quite lengthy and so we only provide typical leading-order terms:

$$R_{SOI,\text{eff}}^{(4,\text{corr})\perp} = R_{SOI,\text{eff}}^{(4)\perp} {}^{-1}\left(\frac{-i\overline{\Delta}}{m_e}\right)$$

$$\left[\sum_{\substack{n_1'\in D_0, \\ n_2\in D_{-1}, \\ n_5\in D_2}} \frac{\sqrt{\tfrac{3}{8}}\lambda_{v,n_5}\Delta_{so}^{(n_1',n_2)}\left(e\zeta_{n_2 v}\right) P_{n_5 n_1'}}{(E_{VBM}-E_{n_2,\uparrow,K})(E_{VBM}-E_{n_1',\downarrow,K})(E_{VBM}-E_{n_5,\downarrow,K})} + \ldots \right], \tag{C33}$$

It can be verified that $R_{SOI,eff}^{(4,corr)\perp}$ are real, with the following orders of magnitude

$$R_{SOI,eff}^{(4,corr)\perp} = O\left[\frac{\bar{\Delta}}{\Delta_{so}} \bar{\lambda}^{(IR-flip)}\right]. \tag{C34}$$

### $g_{eff}^{(3,corr)//}$ in the in-plane Zeeman effect

$$g_{eff}^{(3,corr)//} = -i\frac{g_e\hbar\bar{\Delta}}{2m_e a g_{eff}^{//}} \sum_{n\in D_0} \lambda_{v,n} P_{nv} \left[\frac{1}{(E_{VBM}-E_{n,\downarrow,K})(E_{VBM}-E_{v,\downarrow,K})} + \frac{1}{(E_{VBM}-E_{n,\uparrow,K})(E_{VBM}-E_{v,\downarrow,K})}\right] \tag{C35}$$

It can be verified that $g_{eff}^{(3,corr)//}$ is real, with

$$g_{eff}^{(3,corr)//} = O\left[\bar{\lambda}^{(IR-flip)}\right]. \tag{C36}$$

## APPENDIX D
## ELASTIC SCATTERING

We write the bare Hamiltonian equation

$$\begin{aligned}H\Phi(\vec{r}) &= E\Phi(\vec{r}), \\ H &= H_{etc} + U_{elastic}, \\ \Phi(r) &= \sum_{n,s,\tau} F_{G,s,\tau}^{(n)}(\vec{r})\langle \vec{r}|\Psi_G^{(n)},s,\tau\rangle,\end{aligned} \tag{D1}$$

where the presence of elastic scattering potential energy $U_{elastic}$ is explicitly shown in the Hamiltonian $H$, $H_{etc}$ = Hamiltonian excluding $U_{elastic}$, $\Phi(\vec{r})$ = total wave function, $F_{G,s,\tau}^{(n)}(\vec{r})$ = envelop function, $|\Psi_G^{(n)},s,\tau\rangle$ = band edge state, $n$ = band index, $G$ = irreducible representation (IR) index, $s$ = spin index, and $\tau$ = valley index. Following the standard effective-mass theory, [65]

$$EF_{G,s,\tau}^{(n)}(\vec{R}) \approx N_{cell}\int_{\substack{\text{unit cell}\\ \text{at }\vec{R}}} d\vec{r}\langle\vec{r}|\Psi_G^{(n)},s,\tau\rangle^*\left(H_{etc}+U_{elastic}(\vec{r})\right)\Phi(\vec{r}), \tag{D2}$$

where $N_{cell}$ = total unit cell number, and $\vec{R}$ = lattice vector. $\vec{R}$ appears as the argument in envelop function to indicate that the envelop function is defined with a "unit-cell scale" resolution.

The main task here is to evaluate in Eqn. (D2) the potential energy part and derive the valley-mixing term $U_{K\leftrightarrow K'}$ entering bare models. This is done as follows. We express $\langle\vec{r}|\Psi_G^{(n)},s,\tau\rangle$ as a linear combination of atomic orbitals (or Wannier orbitals):

$$\langle\vec{r}|\Psi_G^{(n)},s,\tau\rangle = \frac{1}{\sqrt{N_{cell}}}\sum_{\vec{R}} e^{i\tau\vec{K}\cdot\vec{R}}\langle\vec{r}-\vec{R}|\Xi_G^{(n)},s,\tau\rangle, \tag{D3}$$

where $\langle\vec{r}-\vec{R}|\Xi_G^{(n)},s,\tau\rangle$ is the corresponding atomic orbital at $\vec{R}$. Substitution of Eqn. (D3) into Eqn. (D2) yields the potential energy part

$$\begin{aligned}&N_{cell}\int_{\substack{\text{unit cell}\\ \text{at }\vec{R}}} d\vec{r}\langle\vec{r}|\Psi_G^{(n)},s,\tau\rangle^* U_{elastic}(\vec{r})\Phi(\vec{r}) \\ &\approx U_{elastic}(\vec{R})F_{G,s,\tau}^{(n)}(\vec{R}) \\ &+ \sum_l e^{-2i\tau\vec{K}\cdot\vec{R}}\lambda_{n,l}^{\tau,-\tau} U_{elastic}^{(derived)}(\vec{R};G,G')F_{G',s,-\tau}^{(l)}(\vec{R}).\end{aligned} \tag{D4}$$

The first term comes from the integral involving states in the same valley, and gives the ordinary, valley-conserving potential energy $U_{diag}$ in the bare model. The second term comes from the integral involving states of opposite valleys, and gives the inter-valley coupling $U_{K\leftrightarrow K'}$ in the bare model, with

$$\begin{aligned}&\lambda_{n,l}^{\tau,-\tau} U_{elastic}^{(derived)}(\vec{R};G,G') \\ &= \int_{\substack{\text{unit cell}\\ \text{at }\vec{R}}} d\vec{r}\langle\vec{r}-\vec{R}|\Xi_G^{(n)},s,\tau\rangle^* U_{elastic}(\vec{r})\langle\vec{r}-\vec{R}|\Xi_{G'}^{(l)},s,-\tau\rangle.\end{aligned} \tag{D5}$$

$\lambda_{n,l}^{\tau,-\tau}$ is a dimensionless strength parameter for the coupling between states with band indices $n$ and $l$, and $U_{elastic}^{(derived)}$ is a potential energy function derived from $U_{elastic}$, which will be specified below. In deriving (D4), we have made a few approximations typically entering the effective-mass theory, for example, the slowly varying approximation for both $U_{elastic}(\vec{r})$ and $F_{G',s',\tau'}^{(l)}(\vec{r})$ on the unit-cell scale; and the "same-site" approximation - the integral vanishes except for orbitals on the same site.

As examples, $\lambda_{n,l}^{\tau,-\tau}$ and $U_{elastic}^{(derived)}$ are given below in a few cases of interest.

(i) In the case of a bulk with dilute, random distribution of identical, short-range impurities on the M-sublattice,

$$\begin{aligned}&\lambda_{n,l}^{\tau,-\tau} \\ &\approx \int_{\text{unit cell}} d\vec{r}\langle\vec{r}|\Xi_G^{(n)},s,\tau\rangle^*\frac{v_{impurity}(\vec{r})}{v_{impurity}(0)}\langle\vec{r}|\Xi_{G'}^{(l)},s,-\tau\rangle, \\ &U_{elastic}^{(derived)}(\vec{R};G,G') = U_{elastic}(\vec{R}).\end{aligned} \tag{D6}$$

(ii) In the case of quantum structures, we write

$$U_{elastic}(\vec{r})$$
$$= U_{elastic}(\vec{R}) + \vec{\nabla}_{\vec{r}} U_{elastic}(\vec{r})\Big|_{\vec{R}} \cdot (\vec{r} - \vec{R})$$
$$+ \frac{1}{2} \sum_{j,h} \frac{\partial^2 U_{elastic}(\vec{r})}{\partial R_j \partial R_h}\Big|_{\vec{R}} (\vec{r} - \vec{R})_j (\vec{r} - \vec{R})_h + \ldots$$
(D7)

Then, for $G = G'$, we have

$$\lambda_{n,l}^{\tau,-\tau} \approx \int_{\text{unit cell}} d\vec{r} \langle \vec{r} | \Xi_G^{(n)}, s, \tau \rangle^* \langle \vec{r} | \Xi_G^{(l)}, s, -\tau \rangle,$$
$$U_{elastic}^{(derived)}(\vec{R}; G, G) \approx U_{elastic}(\vec{R}),$$
(D8)

in the leading order. For $G \neq G'$, we have

$$\lambda_{n,l}^{\tau,-\tau} \approx \frac{1}{2a} \int_{\text{unit cell}} d\vec{r} \langle \vec{r} | \Xi_G^{(n)}, s, \tau \rangle^* r_{\text{sgn}(G,G')} \langle \vec{r} | \Xi_{G'}^{(l)}, s, -\tau \rangle,$$
$$U_{elastic}^{(derived)}(\vec{R}; G, G') \approx a \left[ \vec{\nabla}_{\vec{r}} U_{elastic}(\vec{r})\Big|_{\vec{R}} \right]_{\text{sgn}(G',G)},$$
(D9)

in the leading order. Above, $\text{sgn}(G, G') = -\text{sgn}(G',G) = +$, for $(G, G') = (D_0, D_2)$, $(D_2, D_{-2})$, and $(D_{-1}, D_1)$, and $\text{sgn}(G, G') = 0$ otherwise. $r_0 = 0$, $r_\pm = x \pm iy$, and $\left[\vec{\nabla}_{\vec{r}} U_{elastic}(\vec{r})\Big|_{\vec{R}}\right]_\pm = (\partial_x \pm i\partial_y) U_{elastic}(\vec{r})\Big|_{\vec{R}}$.

It can be shown that $\lambda_{n,l}^{\tau,-\tau} = \lambda_{l,n}^{\tau,-\tau}$ due to the $T$-symmetry, and that $\lambda_{n,l}^{\tau,-\tau}$ is real due to $M_y$ and $T$. In the main context, we denote $\lambda_{n,l}^{\tau,-\tau}$ as $\lambda_{n,l}$. Eqns. (D8) and (D9) describe "IR-diagonal" and "IR-flip" scattering, respectively. In the deep tight-binding regime with extremely narrow atomic orbitals, we have

$$\lambda_{n,l} \leq O(1), \text{ for } G = G';$$
$$\lambda_{n,l} \leq O(a_{TB}/a), \text{ for } G \neq G'.$$
(D10)

where $a_{TB}$ is the orbital size. This indicates that the "IR-diagonal" scattering dominates over the "IR-flip" one in the limit where $a_{TB} \ll a$. In view of such limiting behavior, we divide the quantum paths in our work into IR-diagonal ones (**Class A**) and IR-flip ones (**Class B**), and take, in $H_{eff}$, **Class A** derived Hamiltonian terms to be primary and **Class B** derived terms to be corrections, in the case where both types of paths make contributions to $H_{eff}$ at the same order of perturbation theory.

In a general scenario, the inter-valley scattering may occur at a heterostructure boundary, where band offset-induced potential discontinuities generally differ in strength for different bands. In such a case, for an inter-band valley-flipping scattering, the corresponding $U_{elastic}$ should then be dependent on involved band indices. The present formalism can easily accommodate such dependence by taking the strength $\lambda_{n,l}$ as an empirical parameter. The same generalization applies to the valley-conserving term, namely, $U_{diag}$, in the bare model, where a relative potential strength "$\eta_n$" is assigned to each band, as done in the main text.

† Corresponding author. Email: yswu@ee.nthu.edu.tw

## REFERENCES

[1] M. I. D'yakonov and V. I. Perel', Phys. Lett. **35A**, 459 (1971).
[2] M. I. D'yakonov and V. I. Perel', JETP Lett. **13**, 467 (1971).
[3] J. E. Hirsch, Phys. Rev. Lett. **83**, 1834 (1999).
[4] C. L. Kane and E. J. Mele, Phys. Rev. Lett. **95**, 146802 (2005).
[5] B. A. Bernevig and S.-C. Zhang, Phys. Rev. Lett. **96**, 106802 (2006).
[6] M. N. Baibich, J. M. Broto, A. Fert, F. Nguyen Van Dau, F. Petroff, P. Etienne, G. Creuzet, A. Friederich, and J. Chazelas, Phys. Rev. Lett. **61**, 2472 (1988).
[7] G. Binasch, P. Grunberg, F. Saurenbach, and W. Zinn, Phys. Rev. B **39**, 4828 (1989).
[8] G. H. Jonker and J. H. Van Santen, Physica. **16**, 337 (1950).
[9] S. Jin, T. H. Tiefel1, M. McCormack, R. A. Fastnacht1, R. Ramesh, and L. H. Chen, Science **264**, 413 (1994).
[10] E. I. Rashba, Sov. Phys. Solid State **2**, 1109 (1960).
[11] P. Zeeman, Philosophical Magazine. 5th series. **44**, 55 (1897).
[12] P. Thomas, The Scientific Transactions of the Royal Dublin Society. 2nd series. **6**, 385 (1898).
[13] S. Datta and B. Das, Appl. Phys. Lett. **56**, 665 (1990).
[14] D. Loss and P. Divincenzo, Phys. Rev. A **57**, 120 (1998).
[15] G. Burkard, D. Loss, and D. P. DiVincenzo, Phys. Rev. B **59**, 2070 (1999).
[16] I. Zutić, J. Fabian, and S. Das Sarma, Rev. Mod. Phys. **76**, 323 (2004); and references therein.
[17] S. A. Wolf, D. D. Awschalom, R. A. Buhrman, J. M. Daughton, S. von Molnár, M. L. Roukes, A. Y. Chtchelkanova, D. M. Treger, Science **294**, 1488 (2001).
[18] K. S. Novoselov, A. K. Geim, S. V. Morozov, D. Jiang, M. I. Katsnelson, I. V. Grigorieva, S.V. Dubonos, A. A. Firsov, Nature **438**, 197 (2005).
[19] Y.-B. Zhang, Y.-W. Tan, H. L. Stormer, and P. Kim, Nature **438**, 201 (2005).
[20] K. F. Mak, C.-G Lee, J. Hone, J. Shan, and T. F. Heinz, Phys. Rev. Lett. **105**, 136805 (2010).
[21] Di Xiao, Wang Yao, and Q. Niu, Phys. Rev. Lett. **99**, 236809 (2007).
[22] A. Rycerz, J. Tworzydlo, and C. W. J. Beenakker, Nat. Phys. **3**, 172 (2007).
[23] F. Zhang, A. H. MacDonald, and E. J. Mele, Proc. Natl. Acad. Sci. **110**, 10546 (2013).
[24] W.-K. Tse, Z. Qiao, Y. Yao, A. MacDonald, and Q. Niu, Phys. Rev. B **83**, 155447 (2011).
[25] R. V. Gorbachev, J. C. W. Song, G. L. Yu, A. V. Kretinin, F. Withers, Y. Cao, A. Mishchenko, I. V. Grigorieva, K. S. Novoselov, L. S. Levitov, and A. K. Geim, Science **346**, 448 (2014).
[26] M.-Q. Sui, G.-R. Chen, L.-G. Ma, W.-Y. Shan, D. Tian, K. Watanabe, T. Taniguchi, X.-F. Jin, W. Yao, D. Xiao, and Y.-B. Zhang, Nat. Phys. **11**, 1027 (2015).
[27] Y. Shimazaki, M. Yamamoto, I. V. Borzenets, K. Watanabe, T.

Taniguchi and S. Tarucha, Nature **11**, 1032 (2015).
[28] K. F. Mak, K. L. McGill, J. Park, and P. L. McEuen, Science **334**, 1489 (2014).
[29] D. Xiao, G.-B. Liu, W. Feng, X. Xu, and W. Yao, Phys. Rev. Lett. **108**, 196802 (2012).
[30] G. Y. Wu, N.-Y. Lue, and L. Chang, Phys. Rev. B **84**, 195463 (2011).
[31] T.-C. Hsieh, M.-Y. Chou, and G. Y. Wu, Phys. Rev. M **2**, 034003 (2018).
[32] H. Zeng, J. Dai, W. Yao, D. Xiao, and X. Cui, Nat. Nanotech. **7**, 490 (2012).
[33] K. F. Mak, K. He, J. Shan, and T. F. Heinz, Nat. Nanotech. **7**, 494 (2012).
[34] T. Cao, G. Wang, W.-P. Han, H.-Q. Ye, C.-R. Zhu, J.-R. Shi, Q. Niu, P.-H. Tan, E.-G. Wang, B.-L. Liu, and J. Feng, Nat. Commun. **3**, 887 (2012).
[35] A. M. Jones, H. Yu, N. J. Ghimire, S.-F. Wu, G. Aivazian, J. S. Ross, B. Zhao, J.-Q. Yan, D. G. Mandrus, D. Xiao, W. Yao, and X.-D. Xu, Nat. Nanotech. **8**, 634 (2013).
[36] D. Grundler, Phys. Rev. Lett. **84**, 6074 (2000).
[37] A. M. Gilbertson, M. Fearn, J. H. Jefferson, B. N. Murdin, P. D. Buckle, and L. F. Cohen, Phys. Rev. B **77**, 165335 (2008).
[38] P.-J. Chuang, S.-C. Ho, L. W. Smith, F. Sfigakis, M. Pepper, C.-H. Chen, J.-C. Fan, J. P. Griffiths, I. Farrer, H. E. Beere, G. A. C. Jones, D. A. Ritchie, and T.-M. Chen, Nat. Nanotech. **10**, 35 (2015).
[39] G.-B. Liu, W.-Y. Shan, Y. Yao, W. Yao, and D. Xiao, Phys. Rev. B **88**, 085433 (2013).
[40] H. Rostami, A. G. Moghaddam, and R. Asgari, Phys. Rev. B **88**, 085440 (2013).
[41] A. Kormányos, V. Zólyomi, N. D. Drummond, and G. Burkard, Phys. Rev. X **4**, 011034 (2014).
[42] K. V. Shanavas and S. Satpathy, Phys. Rev. B **91**, 235145 (2015).
[43] G Széchenyi, L Chirolli, and A Pályi, 2D Mater. **5** 035004 (2018).
[44] A. Kormányos, V. Zólyomi, N. D. Drummond, P. Rakyta, G. Burkard, and V. I. Fal'ko, Phys. Rev. B **88**, 045416 (2013)
[45] W. Huang, X. Luo, C.- K. Gan, S.-Y. Quek, and G.-C. Liang, Phys. Chem. **16**, 10866 (2014).
[46] A. Kormányos, G. Burkard, M. Gmitra, J. Fabian, V. Zólyomi, N. D Drummond, and Vladimir Fal'ko, 2D Mater. **2**, 049501 (2015).
[47] S. Fang, R. K. Defo, S. N. Shirodkar, S. Lieu, G. A. Tritsaris, and E. Kaxiras, Phys. Rev. B **92**, 205108 (2015).
[48] X.-X. Song, D. Liu, V. Mosallanejad, J. You, T.-Y. Han, D.-T. Chen, H.-O. Li, G. Cao, M. Xiao, G.-C. Guo, and G.-P. Guo, Nanoscale **7**, 16867 (2015).
[49] Z.-Z. Zhang, X.-X. Song, G. Luo, G.-W. Deng, V. Mosallanejad, T. Taniguchi, K. Watanabe, H.-O. Li, G. Cao, G.-C. Guo, F. Nori, and G.-P. Guo, Sci. Adv. **3**, e1701699 (2017).
[50] K. Wang, K. D. Greve, L. A. Jauregui, A. Sushko, A. High, Y. Zhou, G. Scuri, T. Taniguchi, K. Watanabe, M. D. Lukin, H. Park, and P. Kim, Nat. Nanotech. **13,** 128 (2018).
[51] R. Pisoni, Z. Lei, P. Back, M. Eich, H. Overweg, Y. Lee, K. Watanabe, T. Taniguchi, T. Ihn, and K. Ensslin, Appl. Phys. Lett. **112** 123101 (2018).
[52] R. Pisoni, Y.-J. Lee, H. Overweg, M. Eich, P. Simonet, K.Watanabe, T. Taniguchi, R. Gorbachev, T. Ihn, and K. Ensslin, Nano Lett. **17**, 5008 (2017).
[53] N. Rohling and G. Burkard, New J. Phys. **14** 083008 (2012).
[54] N. Rohling, M. Russ, and G. Burkard, Phys. Rev. Lett. **113**, 176801 (2014).
[55] Y. Wu, Q.-J. Tong, G.-B. Liu, H.-Y. Yu, and W. Yao, Phys. Rev. B **93**, 045313 (2016).
[56] J. Pawłowski, D. Żebrowski, and S. Bednarek, Phys. Rev. B **97**, 155412 (2018).
[57] J. Kim, C. Jin, B. Chen, H. Cai, T. Zhao, P. Lee, S. Kahn, K. Watanabe, T. Taniguchi, S. Tongay, M. F. Crommie, and F. Wang, Sci. Adv. **3**, e1700518 (2017).
[58] P. Dey, L.-Y. Yang, C. Robert, G. Wang, B. Urbaszek, X. Marie, and S. A. Crooker, Phys. Rev. Lett. **119**, 137401 (2017).
[59] C.-H Jin, J.-H Kim, M. I. B Utama, E. C. Regan, H. Kleemann, H. Cai, Y.-X. Shen, M. J. Shinner, A. Sengupta, K. Watanabe, T. Taniguchi, S. Tongay, A. Zett, F. Wang, Science **360**, 893 (2018).
[60] C. J. Ciccarino, T. Christensen, R. Sundararaman, and P. Narang, Nano Lett. **18**, 5709 (2018).
[61] G.-B. Liu, H. Pang, Y. Yao, and W. Yao, New J. Phys. **16**, 105011 (2014).
[62] H. Rostami, R. Asgari, and F. Guinea, J. Phys.: Condens. Matter **28** 495001 (2016).
[63] J. Kang, S. Tongay, J. Zhou, J. Li, and J. Wu, Appl. Phys. Lett. **102**, 012111 (2013).
[64] M.-Y. Li, Y.-M. Shi, C.-C. Cheng, L.-S. Lu, Y.-C. Lin, H.-L. Tang, M.-L. Tsai, C.-W. Chu, K.-H. Wei, J.-H. He, W.-H. Chang, K. Suenaga, and L.-J. Li, Science **349**, 524 (2015).
[65] J. M. Luttinger and W. Kohn, Phys. Rev. **97,** 869 (1955).
[66] M. Koperski, M. R. Molas, A. Arora, K. Nogajewski, A. O. Slobodeniuk, C. Faugeras, and M. Potemski, Nanophotonics **6**, 1289 (2017).
[67] M. Born and V. A. Fock, Z. Phys. A **51**, 165 (1928).
[68] I. I. Rabi, N. F. Ramsey, and J. Schwinger, Rev. Mod. Phys. **26**, 167 (1954).
[69] E. V. Castro, K. S. Novoselov, S. V. Morozov, N. M. R. Peres, J. M. B. Lopes dos Santos, J. Nilsson, F. Guinea, A. K. Geim, and A. H. Castro Neto, Phys. Rev. Lett. 99, 216802 (2007).
[70] H. Overweg, H. Eggimann, X. Chen, S. Slizovskiy, M. Eich, R. Pisoni, Y.-G. Lee, P. Rickhaus, K. Watanabe, T. Taniguchi, V. Fal'ko, T. Ihn, K. Ensslin, Nano Lett. **18**, 553 (2018).
[71] V. N. Golovach, M. Borhani, and D. Loss, Phys. Rev. B **74**, 165319 (2006).
[72] D. P. DiVincenzo, Fortschr. Phys. **48**, 771 (2000).
[73] J. M. Elzerman, R. Hanson, L. H. Willems van Beveren, B. Witkamp, L. M. K. Vandersypen, and L. P. Kouwenhoven, Nat, **430**, 431 (2004).
[74] R. Hanson, L. H. Willems van Beveren, I. T. Vink, J. M. Elzerman, W. J. M. Naber, F. H. L. Koppens, L. P. Kouwenhoven, and L. M. K. Vandersypen, Phys. Rev. Lett **94**, 196802 (2005).
[75] Z. Y. Zhu, Y. C. Cheng, and U. Schwingenschlögl, Phys. Rev. B **84**, 153402 (2011).
[76] J. R. Schrieffer and P. A. Wolff, Phys. Rev. **149**, 491 (1966).
[77] R. Winkler, Appendix B, *Spin-Orbit Coupling Effects in Two-Dimensional Electron and Hole Systems* (Springer, Berlin, 2003).